\documentclass[lettersize,journal]{IEEEtran}
\usepackage{amsmath, amsfonts, amssymb}
\usepackage{algorithmic}
\usepackage{algorithm}
\usepackage{array}
\usepackage{pifont}
\usepackage[caption=false,font=normalsize,labelfont=scriptsize,textfont=scriptsize]{subfig}
\usepackage{textcomp}
\usepackage{stfloats}
\usepackage{url}
\usepackage{enumitem}
\usepackage{verbatim}
\usepackage{graphicx}
\usepackage{makecell}
\usepackage{cite}
\newtheorem{remark}{\textbf{Remark}}
\newtheorem{proposition}{\textbf{Proposition}}
\begin{document}
\title{Flexible Intelligent Metasurfaces for\\Enhancing MIMO Communications}
\author{Jiancheng An,~\IEEEmembership{Member,~IEEE}, Zhu Han,~\IEEEmembership{Fellow,~IEEE}, Dusit Niyato,~\IEEEmembership{Fellow,~IEEE},\\M\'erouane Debbah,~\IEEEmembership{Fellow,~IEEE}, Chau Yuen,~\IEEEmembership{Fellow,~IEEE}, and Lajos Hanzo,~\IEEEmembership{Life Fellow,~IEEE}
\thanks{This research is supported in part by the Agency for Science, Technology and Research M22L1b0110, Ministry of Education Tier2 T2EP50124-0032, Infocomm Media Development Authority FCP-NTU-RG-2024-025. This work is partially supported by NSF ECCS-2302469, CMMI-2222810, Toyota. Amazon and Japan Science and Technology Agency (JST) Adopting Sustainable Partnerships for Innovative Research Ecosystem (ASPIRE) JPMJAP2326. This research is supported by the National Research Foundation, Singapore and Infocomm Media Development Authority under its Future Communications Research \& Development Programme (FCP-NTU-RG-2022-010 and FCP-ASTAR-TG-2022-003), Singapore Ministry of Education (MOE) Tier 1 (RG87/22 and RG24/24), the NTU Centre for Computational Technologies in Finance (NTU-CCTF), and the RIE2025 Industry Alignment Fund - Industry Collaboration Projects (IAF-ICP) (Award I2301E0026), administered by A\*STAR. L. Hanzo would like to acknowledge the financial support of the Engineering and Physical Sciences Research Council (EPSRC) projects under grant EP/Y037243/1, EP/W016605/1, EP/X01228X/1, EP/Y026721/1, EP/W032635/1, EP/Y037243/1 and EP/X04047X/1 as well as of the European Research Council's Advanced Fellow Grant QuantCom (Grant No. 789028). \emph{(Corresponding author: Chau Yuen.)}}
\thanks{J. An and C. Yuen are with the School of Electrical and Electronics Engineering, Nanyang Technological University, Singapore 639798 (e-mail: jiancheng.an@ntu.edu.sg, chau.yuen@ntu.edu.sg).}
\thanks{Z. Han is with the Department of Electrical and Computer Engineering at the University of Houston, Houston, TX 77004 USA, and also with the Department of Computer Science and Engineering, Kyung Hee University, Seoul, South Korea, 446-701 (e-mail: hanzhu22@gmail.com).}
\thanks{D. Niyato is with the College of Computing and Data Science, Nanyang Technological University, Singapore (e-mail: dniyato@ntu.edu.sg).}
\thanks{M. Debbah is with KU 6G Research Center, Department of Computer and Information Engineering, Khalifa University, Abu Dhabi 127788, UAE (e-mail: merouane.debbah@ku.ac.ae) and also with CentraleSupelec, University Paris-Saclay, 91192 Gif-sur-Yvette, France.}
\thanks{L. Hanzo is with the School of Electronics and Computer Science, University of Southampton, Southampton SO17 1BJ, U.K. (e-mail: lh@ecs.soton.ac.uk).}\vspace{-1cm}}
\markboth{IEEE TCOM}{IEEE TCOM}
\maketitle
\begin{abstract}
Flexible intelligent metasurfaces (FIMs) show great potential for improving the wireless network capacity in an energy-efficient manner. An FIM is a soft array consisting of several low-cost radiating elements. Each element can independently emit electromagnetic signals, while flexibly adjusting its position even perpendicularly to the overall surface to `morph' its 3D shape. More explicitly, compared to a conventional rigid antenna array, an FIM is capable of finding an optimal 3D surface shape that provides improved signal quality. In this paper, we study point-to-point multiple-input multiple-output (MIMO) communications between a pair of FIMs. In order to characterize the capacity limits of FIM-aided MIMO transmissions over frequency-flat fading channels, we formulate a transmit optimization problem for maximizing the MIMO channel capacity by jointly optimizing the 3D surface shapes of the transmitting and receiving FIMs as well as the MIMO transmit covariance matrix, subject to the total transmit power constraint and to the maximum perpendicular morphing range of the FIM. To solve this problem, we develop an efficient block coordinate descent (BCD) algorithm. The BCD algorithm iteratively updates the 3D surface shapes of the FIMs and the transmit covariance matrix, while keeping the other fixed, to find a locally optimal solution. Numerical results verify that FIMs can achieve higher MIMO capacity than that of the conventional rigid arrays. In particular, the MIMO channel capacity can be doubled by the proposed BCD algorithm under some setups.
\end{abstract}


\section{Introduction}
\subsection{Background}
\IEEEPARstart{T}{he} sixth generation (6G) of mobile networks promise to enable innovative applications such as brain-computer interfaces, high-fidelity holographic videos, and soft robotics \cite{SCIS_2021_You_Towards}. As these emerging mobile applications proliferate explosively, there is an ever-increasing demand for powerful network capabilities in terms of increased data rates, ubiquitous and ultra-reliable device connectivity, extremely low latency, and scalable power dissipation \cite{Proc_2021_Tataria_6G}. Unfortunately, the existing network infrastructures cannot fulfill these stringent demands. This motivates wireless researchers to develop radically new wireless technologies \cite{JSAC_2020_Zhang_Prospective}. In the fifth generation (5G) of wireless networks, massive multiple-input multiple-output (MIMO) has become a core technology \cite{CM_2016_Bjornson_Massive}. As 6G imposes higher capacity requirements, various radical MIMO technologies have emerged in recent years for further enhancing the spectral efficiency of wireless networks \cite{arXiv_2023_An_Stacked_mag, arXiv_2025_An_Emerging}. In the following, we briefly overview the major evolutionary directions of the MIMO technology from four aspects: array aperture, element density, architecture, and array flexibility.
\subsubsection{MIMOs having Large Aperture}
Classic MIMO technologies aim to attain diversity and/or spatial multiplexing gain by utilizing multiple antennas at both the source and destination. By leveraging the channel state information (CSI) available at transceivers and applying appropriate precoding/combining techniques, higher data rates can be achieved within each channel use \cite{JSAC_2003_Goldsmith_Capacity, TIT_2003_Zheng_Diversity}. In 5G networks, base stations (BSs) typically use large-scale antenna arrays having more than $100$ antennas \cite{TWC_2010_Marzetta_Noncooperative}. Furthermore, by leveraging a large number of distributed access points for simultaneously serving multiple users within the same time/frequency resource blocks, cell-free massive MIMO systems achieve considerably better performance than ultra-dense networks \cite{TWC_2017_Ngo_Cell}, particularly in the presence of shadow fading.

As expected, extremely large-scale antenna arrays (ELAAs) equipped with thousands of antennas will become a common feature of next-generation communication systems \cite{TWC_2022_Lu_Communicating}. Since the antenna aperture can be very large, electromagnetic radiation propagation will likely occur in the near-field Fresnel region of the ELAA. As a result, near-field communications (NFC) will become an emerging technology in 6G wireless networks \cite{arXiv_2023_An_Toward}. To accurately portray the channel characteristics of near-field wireless propagation, spherical wavefront-based models are required. This allows NFC to harvest the full spatial multiplexing gain even under strong line-of-sight (LoS) propagation conditions \cite{arXiv_2023_An_Toward, CM_2023_Cui_Near}.

\subsubsection{MIMOs using Dense Array}
In practical MIMO communications, a major challenge, especially for mobile terminals, is how to pack a large number of antenna elements into a limited physical area. A promising solution that is gaining popularity is the metasurface technology \cite{TCOM_2022_An_Low, LSA_2014_Cui_Coding}. Specifically, a metasurface is an artificially engineered planar structure comprised of a large number of sub-wavelength metallic or dielectric scattering particles \cite{LSA_2014_Cui_Coding}, each of which is capable of independently radiating or scattering electromagnetic waves as desired \cite{JSAC_2020_Renzo_Smart}. Incorporating near-passive metasurfaces into wireless networks, on the one hand, has great potential for reshaping the wireless propagation environments \cite{TCOM_2022_An_Low, WC_2022_An_Codebook}. By controlling the phase shift and/or amplitude attenuation of the signal incident upon each meta-atom in real-time, the wireless channels between transmitters and receivers can be modified to achieve more favorable channel conditions.

On the other hand, metasurfaces can also be used as reconfigurable antenna arrays thanks to their tunability and programmability. For instance, an appealing dynamic metasurface antenna (DMA) technology was proposed in \cite{WC_2021_Shlezinger_Dynamic} for next-generation MIMO transceivers. DMAs offer reconfigurable beamforming capabilities with reduced power consumption and hardware costs. Moreover, a reconfigurable holographic surface (RHS) composed of numerous metamaterial radiation elements was developed in \cite{WC_2021_Deng_Reconfigurable, CM_2022_Zhang_Intelligent, JSAC_2022_Zhang_Holographic}. The RHS serves as an ultra-thin and lightweight surface antenna integrated with transceivers, which is capable of generating the desired beam patterns by leveraging holographic principles. Since metasurfaces having dense arrays enable electromagnetic operations at an unprecedented level of resolution \cite{CM_2021_Dardari_Holographic}, the concept of holographic MIMO communications has recently emerged for characterizing the metasurface's performance limits by modeling it as a near-continuous array relying on a massive number of infinitesimally small antennas \cite{CL_2023_An_A1, TWC_2022_Pizzo_Fourier, CL_2023_An_A2, CL_2023_An_A3}.

\begin{table*}[!t]
\renewcommand\arraystretch{1.5}
\centering
\caption{Four typical MIMO evolution directions.}
\begin{tabular}{l||c|l|l}
\hline
 & Space & Common concepts & Features \\ \hline
Large aperture & 2D & \makecell[l]{MIMO \cite{BOOK_2005_Tse_Fundamentals}, Massive MIMO \cite{CM_2016_Bjornson_Massive},\\ELAA \cite{TWC_2022_Lu_Communicating}, NFC \cite{arXiv_2023_An_Toward, CM_2023_Cui_Near}} & Deploying multiple antennas to achieve spatial multiplexing and diversity gains. \\ \hline
Dense placement & 2D & \makecell[l]{Holographic MIMO \cite{CL_2023_An_A1, TWC_2022_Pizzo_Fourier, CL_2023_An_A2, CM_2021_Dardari_Holographic, CL_2023_An_A3},\\LIS\cite{TSP_2018_Hu_Beyond, TWC_2020_Jung_Performance}, RIS \cite{JSAC_2020_Renzo_Smart}, RHS \cite{WC_2021_Deng_Reconfigurable}} & Leveraging dense arrays to achieve the full array gain. \\ \hline
Deep architecture & 3D & \makecell[l]{Hybrid MIMO \cite{JSTSP_2016_Heath_An, TWC_2014_Ayach_Spatially},\\SIM \cite{arXiv_2023_An_Stacked_mag, arXiv_2023_An_Stacked_DOA, JSAC_2023_An_Stacked}} & Stacking multiple programmable metasurfaces to perform analog computation. \\ \hline
Flexible array & 3D & \makecell[l]{FAS \cite{TWC_2022_Wong_Fluid}, MAS \cite{TWC_2023_Ma_MIMO}, FIM $\star$} & Morphing the surface shape of array substrate to attain selection diversity gain. \\ \hline
\end{tabular}\vspace{-0.5cm}
\end{table*}
\subsubsection{MIMOs Relying on Deep Architecture}
Another challenge for the application of MIMO is that the conventional fully digital architecture requires a costly dedicated radio frequency (RF) chain for each active antenna element. To address this issue, the hybrid MIMO architecture has been proposed and widely adopted in 5G communication systems \cite{TWC_2014_Ayach_Spatially}. Specifically, a low-dimensional digital beamformer is followed by an analog beamformer relying on analog phase shifters \cite{JSTSP_2016_Sohrabi_Hybrid}. Hybrid beamforming can be implemented using a reduced number of RF chains, although some practical tuning constraints of the phase shifters have to be considered.

Additionally, advances in three-dimensional (3D) metasurface technology have made analog signal processing in the electromagnetic domain more attractive, catalyzing the emerging technology of stacked intelligent metasurfaces (SIM) \cite{WCL_2025_Huang_Stacked, WCL_2024_Niu_Stacked, TCCN_2025_Hao_Multi, ICC_2024_An_Stacked, WCL_2024_Lin_Stacked}. A SIM is created by stacking multiple programmable metasurface layers. By appropriately configuring the meta-atoms, SIMs can carry out various advanced computation and signal processing tasks, such as MIMO precoding \cite{JSAC_2023_An_Stacked} and direction-of-arrival (DOA) estimation \cite{arXiv_2023_An_Stacked_DOA}, as the electromagnetic waves propagate through an optimized SIM at the speed of light \cite{NE_2022_Liu_A}. By utilizing SIMs to perform MIMO precoding and combining \cite{arXiv_2023_An_Stacked_DOA, JSAC_2023_An_Stacked}, each data stream can be directly radiated and recovered from its corresponding transmit and receive ports without requiring an excessive number of RF chains and high-resolution analog-to-digital converters \cite{arXiv_2023_An_Stacked_mag}. Accordingly, the energy consumption, hardware cost, transceiver complexity, and processing delay are substantially reduced compared to traditional full-digital and hybrid MIMO architectures \cite{arXiv_2023_An_Stacked, arXiv_2024_Liu_Stackedmag}.
\subsubsection{MIMOs Based on Flexible Array}
It is noted that all of the aforementioned MIMO technologies are based on rigid antenna arrays (RAAs) with fixed element positions, which may lead to weak channel quality when deep fading occurs. However, strategically changing the physical position of each antenna element may guarantee that the multiple signal copies impinging from different paths add constructively at the antenna array, thus boosting the received signal power \cite{BOOK_2005_Tse_Fundamentals}. This is particularly beneficial for wireless communications at high millimeter-wave and terahertz frequencies, as the coherence distance is very small. Following this philosophy, a fluid antenna system (FAS) was recently proposed in \cite{TWC_2022_Wong_Fluid} that allows an antenna to be flexibly switched to the position having the strongest signal along a line, thereby reaping extra selection diversity. Furthermore, the authors in \cite{TWC_2023_Ma_MIMO} extended this flexible MIMO concept to point-to-point MIMO communication systems using movable antennas (MASs) driven by electric motors, which significantly improved the capacity compared to conventional MIMO systems having fixed antenna positions.
\begin{figure}[!t]
\centering
\includegraphics[width = 8 cm]{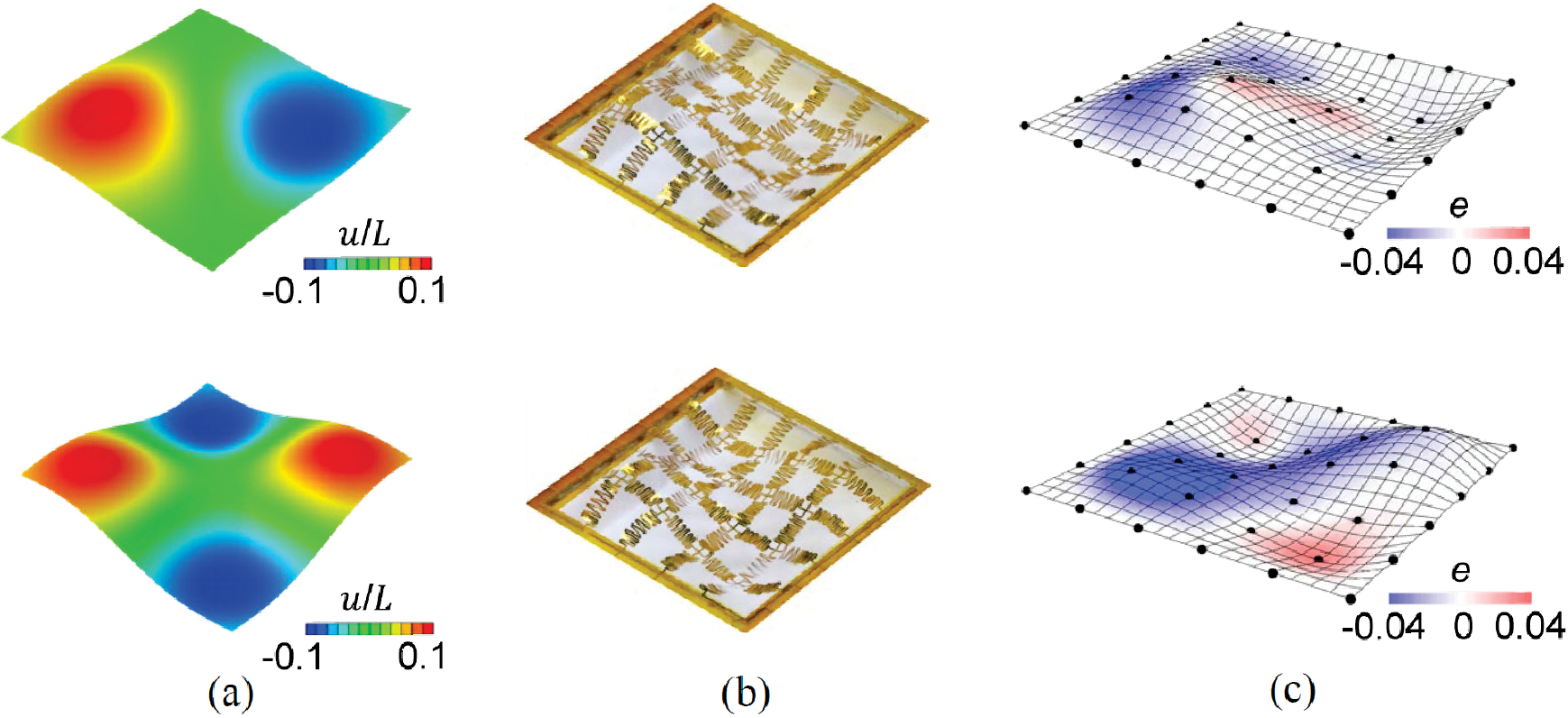}
\caption{Illustration of an existing FIM prototype \cite{Nature_2022_Bai_A}, where (a) target surface shapes; (b) experimental results of morphed surface shape; (c) simulation results of morphed surface shape.}\vspace{-0.5cm}
\label{fig_1}
\end{figure}

Despite the potential of flexible MIMO technologies using FAS or MAS, they also face some limitations. For example, the change from one position to another is restricted by the switching speeds of fluid antennas and electric motors. Additionally, these flexible MIMO concepts cannot be readily combined with metasurface-based principles. Fortunately, recent advances in micro/nanofabrication and the discovery of flexible metamaterials make it possible to create flexible intelligent metasurfaces (FIMs) by depositing dielectric layers onto a soft conformal substrate such as polydimethylsiloxane (PDMS), which have shown great potential for wave manipulation on curved and conformal surfaces \cite{NC_2016_Kamali_Decoupling}. Extensive research has examined the use of FIMs as electromagnetic skins, highlighting their potential to significantly reduce radar cross-section (RCS) \cite{APSIS_2002_Georgakopoulos_HIRF, TAP_2020_Modi_Metasurface, ISAP_2018_Modi_Investigation}. More recently, a programmable FIM was developed, enabling dynamic 3D surface-shape morphing \cite{Nature_2022_Bai_A}. As illustrated in Fig. \ref{fig_1}, this dynamic FIM is composed of a matrix of tiny metallic filaments, which are manipulated by reprogrammable distributed Lorentz forces generated by electrical currents passing through a static magnetic field\footnote{Please refer to \url{https://www.eurekalert.org/multimedia/950133} for a video demonstrating the real-time surface-shape morphing capability of an FIM \cite{Nature_2022_Bai_A}.}. This innovative design endows the FIM with impressive capabilities for rapidly and precisely morphing its surface shape \cite{Nature_2022_Bai_A}. Note that FIMs inherit both the selection gain of flexible MIMOs and the array gain of metasurfaces. Therefore, an FIM is expected to be a promising technology for further improving the spectral and energy efficiencies of next-generation wireless networks \cite{ICC_2025_Jiancheng_Flexible, GLOBECOM_2025_Jiancheng_Downlink, TWC_2025_An_Flexible}.

\subsection{Contributions}
Despite the significant potential, the application of FIMs in wireless networks remains largely unexplored. Against this background, in this paper, we harness FIMs as transceivers in a point-to-point MIMO communication system. The underlying rationale is that by flexibly morphing the 3D surface shapes of the transmitting and receiving FIMs, the system can ensure that the multipath signal components add together constructively at the selected antenna positions, thus achieving improved channel capacity. To characterize the fundamental capacity limit, we consider that perfect CSI is available at both the source and the destination \cite{TWC_2023_Ma_MIMO, JSAC_2023_An_Stacked}. Specifically, our main contributions are summarized as follows:
\begin{itemize}
\item First, we develop the system and channel models for characterizing FIM-aided MIMO communications for transmission over narrowband frequency-flat channels. Based on these models, we formulate an optimization problem for maximizing the MIMO channel capacity by jointly optimizing the transmit signal covariance matrix and the 3D surface shapes of the transmitting and receiving FIMs, subject to realistic constraints on the total transmit power at the source and the maximum perpendicular morphing range of the FIMs at the source and destination.
\item To solve the problem formulated, we propose an efficient block coordinate descent (BCD) algorithm to optimize the transmit covariance matrix and the 3D surface shapes of the transmitting and receiving FIMs in an iterative manner. In each iteration, the optimal transmit covariance matrix is derived in a closed form given the FIM's 3D surface shapes in the previous iteration, while the surface shapes of the FIMs are updated to increase the channel capacity utilizing a customized gradient ascent algorithm. The proposed BCD algorithm is guaranteed to converge to at least a locally optimal solution.
\item Finally, we examine the computational complexity of the proposed BCD algorithm. Extensive numerical results validate the superior performance of the BCD algorithm compared to other benchmark schemes. The results demonstrate that by judiciously morphing the 3D surface shapes of the transmitting and receiving FIMs, the MIMO capacity is doubled under some setups.
\end{itemize}
\begin{table}[!t]
\renewcommand\arraystretch{1.5}
\centering
\scriptsize
\caption{List of major symbols and their meanings}
\begin{tabular}{c|l}
\hline
Symbol & Description \\ \hline
$\varphi_{\textrm{t}}$ ($\varphi_{\textrm{r}}$) & \makecell[l]{The azimuth angle of the normal direction of the transmitting \\ (receiving) FIM} \\ \hline
$\vartheta_{\textrm{t}}$ ($\vartheta_{\textrm{r}}$) & \makecell[l]{The elevation angle of the normal direction of the transmitting\\ (receiving) FIM} \\ \hline
$\varrho_{\textrm{t}}$ ($ \varrho_{\textrm{r}}$) & The spin angle of the transmitting (receiving) FIM \\ \hline
$\mathbf{i}_{\textrm{t}}$ ($\mathbf{i}_{\textrm{r}}$), $\mathbf{j}_{\textrm{t}}$ ($\mathbf{j}_{\textrm{r}}$) & The directions of two sides of the transmitting (receiving) FIM \\ \hline
$\mathbf{k}_{\textrm{t}}$ ($\mathbf{k}_{\textrm{r}} $) & The normal direction of the transmitting (receiving) FIM \\ \hline
$M$, $N$ & Numbers of transmitting and receiving antenna elements \\ \hline
$M_{\textrm{x}}$, $M_{\textrm{y}}$ & Numbers of transmitting antennas in the $\mathbf{i}_{\textrm{t}}$ and $\mathbf{j}_{\textrm{t}}$ directions\\ \hline
$N_{\textrm{x}}$, $N_{\textrm{y}}$ & Numbers of receiving antennas in the $\mathbf{i}_{\textrm{r}}$ and $\mathbf{j}_{\textrm{r}}$ directions\\ \hline
$d_{\textrm{t},\textrm{x}}$, $d_{\textrm{t},\textrm{y}}$ & Spacing between adjacent transmitting antenna elements\\ \hline
$d_{\textrm{r},\textrm{x}}$, $d_{\textrm{r},\textrm{y}}$ & Spacing between adjacent receiving antenna elements\\ \hline
$x_{m}^{\textrm{t}}$, $y_{m}^{\textrm{t}}$ & \makecell[l]{The distances of the $m$-th transmitting element \emph{w.r.t.} the\\reference antenna along the $\mathbf{i}_{\textrm{t}}$ and $\mathbf{j}_{\textrm{t}}$ directions} \\ \hline
$x_{n}^{\textrm{r}}$, $y_{n}^{\textrm{r}}$ & \makecell[l]{The distances of the $n$-th receiving element \emph{w.r.t.} the\\reference antenna along the $\mathbf{i}_{\textrm{r}}$ and $\mathbf{j}_{\textrm{r}}$ directions} \\ \hline
$\zeta_{m} \left ( \xi_{n} \right ) $ & \makecell[l]{The deformation distance of the $m$-th transmitting\\($n$-th receiving) antenna element}\\ \hline
$\boldsymbol{\zeta}$ ($\boldsymbol{\xi} $) & \makecell[l]{The deformation vector of the transmitting (receiving) FIM} \\ \hline
$\mathbf{q}_{m}$, $ \tilde{\mathbf{q}}_{m}$ & \makecell[l]{The positions of the $m$-th transmitting element} \\ \hline
$\mathbf{p}_{n}$, $\tilde{\mathbf{p}}_{n}$ & The positions of the $n$-th receiving element \\ \hline
$\varphi_{l,g}^{\textrm{t}}$ ($ \varphi_{l,g}^{\textrm{r}} $) & \makecell[l]{The azimuth angle of the $g$-th scatter in the $l$-th cluster \emph{w.r.t.} the\\transmitting (receiving) FIM} \\ \hline
$\vartheta_{l,g}^{\textrm{t}}$ ($ \vartheta_{l,g}^{\textrm{r}} $) & \makecell[l]{The elevation angle of the $g$-th scatter in the $l$-th cluster \emph{w.r.t.} the\\transmitting (receiving) FIM} \\ \hline
$\mathbf{o}_{l,g}^{\textrm{t}}$ ($\mathbf{o}_{l,g}^{\textrm{r}}$) & \makecell[l]{Propagation direction of the $g$-th path in the $l$-th cluster \emph{w.r.t.} the\\transmitting (receiving) FIM}\\ \hline
$L$ & Number of scattering clusters\\ \hline
$G$ & Number of propagation paths in each cluster\\ \hline
$\mathbf{t}$, $\mathbf{T}$ & \makecell[l]{The transmit signal vector and corresponding covariance matrix}\\ \hline
\end{tabular}\vspace{-0.5cm}
\end{table}
\subsection{Organization}
The rest of this paper is organized as follows. Section \ref{sec2} introduces the system and channel models of an FIM-aided point-to-point MIMO system, and presents the problem formulation. Section \ref{sec3} proposes an efficient BCD algorithm for solving the problem formulated and also analyzes its convergence and complexity. Section \ref{sec4} presents numerical results to evaluate the performance of the proposed FIM-aided MIMO communication systems. Finally, Section \ref{sec5} concludes this paper.

\subsection{Notation}
Scalars are denoted by italic letters. Vectors and matrices are represented by bold-face lowercase and uppercase letters, respectively. For a complex-valued vector $\mathbf{v}$, $\left \| \mathbf{v} \right \|$ represents its Euclidean norm, and $\textrm{diag}\left ( \mathbf{v} \right )$ represents a diagonal matrix with the elements of $\mathbf{v}$ on its main diagonal, while $\left \langle \mathbf{v}, \mathbf{w} \right \rangle$ denotes the cosine of the angle between vectors $\mathbf{v}$ and $\mathbf{w}$. For a square matrix $\mathbf{S}$, $\mathbf{S}^{-1}$, $\textrm{tr} \left (\mathbf{S} \right )$, and $\det \left (\mathbf{S} \right )$ denote its inverse, trace, and determinant, respectively, while $\mathbf{S}\succeq \mathbf{0}$ means that $\mathbf{S}$ is positive semidefinite. $\textrm{Diag}\left ( \mathbf{S} \right )$ represents a diagonal matrix with the diagonal elements of $\mathbf{S}$ on its main diagonal. For any general matrix $\mathbf{M}$, $\mathbf{M}^{T}$, $\mathbf{M}^{H}$, and $\left [ \mathbf{M} \right ]_{p,q}$ denote its transpose, Hermitian transpose, and the $\left ( p,q \right )$-th element, respectively. Furthermore, $\mathbf{0}_{M \times N}$ represents an $M \times N$ all-zero matrix, while $\mathbf{I}_N$ denotes an $N \times N$ identity matrix. $\mathbb{E}\left ( \cdot \right )$ stands for the statistical expectation, and $\odot $ represents the Hadamard product. $\Im \left ( \cdot \right )$ denotes the imaginary part of a complex number. $\log_{2}\left ( \cdot \right )$ represents the logarithmic function with base $2$, while $\ln\left ( \cdot \right )$ denotes the natural logarithm. $\textrm{mod}\left ( x,y \right )$ returns the remainder from dividing $x$ by $y$. The notation $\left \lfloor \cdot \right \rfloor$ represents the floor function, and $d\left ( \cdot \right )$ represents the differential operator. The gradient of a function $f$ with respect to (\emph{w.r.t.}) $\mathbf{x}$ is denoted by $\nabla_\mathbf{x} f$, while the partial derivative of $g$ \emph{w.r.t.} the variable $x$ is denoted by $\frac{\partial}{\partial x}g$. $\mathbb{R}^{n}$ and $\mathbb{C}^{n}$ represent the $n$-dimensional spaces of real and complex vectors, respectively, while $\mathbb{R}^{m\times n}$ and $\mathbb{C}^{m\times n}$ denote the spaces of $m\times n$ real and complex matrices, respectively. The distribution of a circularly symmetric complex Gaussian (CSCG) random vector with mean vector $\mathbf{v}$ and covariance matrix $\boldsymbol{\Sigma}$ is denoted by $\sim \mathcal{CN}\left ( \mathbf{v},\boldsymbol{\Sigma} \right )$, where $\sim $ means ``distributed as''. $\mathcal{U}\left ( x,y \right )$ represents the uniform distribution over the interval $\left ( x,y \right )$.

\begin{figure}[!t]
\centering
\includegraphics[width = 7 cm]{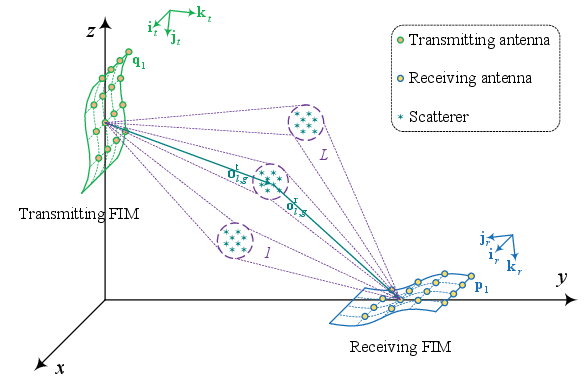}
\caption{Schematic of a point-to-point MIMO system, where an FIM is deployed at the source and another one is at the destination.}\vspace{-0.5cm}
\label{fig_2}
\end{figure}

\section{Communication Model and Problem Formulation}\label{sec2}

\subsection{System Model}
As shown in Fig. \ref{fig_2}, we consider a point-to-point MIMO communication system with a pair of FIMs communicating with each other. A transmitting FIM is deployed at the source, while a receiving FIM is deployed at the destination. The antenna arrays on the two FIMs are modeled as a flexible uniform planar array (UPA). In contrast to conventional MIMO systems, the position of each radiating element on the FIMs can be dynamically adjusted along the direction perpendicular to the surface, with the aid of a controller \cite{Nature_2022_Bai_A}. For simplicity, each element on the FIM is abstracted as a point antenna. Additionally, we consider a general placement of the transmitting and receiving FIMs with arbitrary array orientations and spin angles.
\subsubsection{Transmitting FIM}
Specifically, let $\mathbf{i}_{\textrm{t}}\in \mathbb{R}^{3}$ and $\mathbf{j}_{\textrm{t}}\in \mathbb{R}^{3}$ represent the directions of the two sides of the transmitting FIM, while $\mathbf{k}_{\textrm{t}}\in \mathbb{R}^{3}$ represents the normal direction of the transmitting FIM. Furthermore, let $\varphi_{\textrm{t}}\in \left [ 0,\pi \right )$ and $\vartheta_{\textrm{t}}\in \left [ 0,\pi \right )$ denote the azimuth and elevation angles, respectively, of the normal direction. Hence, $\mathbf{k}_{\textrm{t}}$ can be expressed as
\begin{align}\label{eq1}
 \mathbf{k}_{\textrm{t}} = \left [ \sin\vartheta _{\textrm{t}}\cos\varphi _{\textrm{t}},\sin\vartheta_{\textrm{t}}\sin\varphi _{\textrm{t}},\cos\vartheta _{\textrm{t}} \right ] ^{T}.
\end{align}

Due to the fact that $\mathbf{i}_{\textrm{t}}$, $\mathbf{j}_{\textrm{t}}$, and $\mathbf{k}_{\textrm{t}}$ are orthogonal to each other, $\mathbf{i}_{\textrm{t}}$ and $\mathbf{j}_{\textrm{t}}$ can be derived by
\begin{align}
 \mathbf{i}_{\textrm{t}} =& \left [ \cos\vartheta_{\textrm{t}} \cos\varphi_{\textrm{t}} \cos\varrho_{\textrm{t}}-\sin\varphi_{\textrm{t}} \sin \varrho_{\textrm{t}}, \cos\vartheta_{\textrm{t}} \right. \notag\\
 & \left. \times \sin\varphi_{\textrm{t}} \cos\varrho_{\textrm{t}}+\cos\varphi_{\textrm{t}} \sin \varrho_{\textrm{t}}, -\sin\vartheta_{\textrm{t}} \cos\varrho_{\textrm{t}} \right ]^{T}, \label{eq2}\\
 \mathbf{j}_{\textrm{t}} =& \left [ -\cos\vartheta_{\textrm{t}} \cos\varphi_{\textrm{t}} \sin\varrho_{\textrm{t}}-\sin\varphi_{\textrm{t}} \cos\varrho_{\textrm{t}},-\cos\vartheta_{\textrm{t}} 
 \right. \notag\\
 & \left. \times \sin\varphi_{\textrm{t}} \sin\varrho_{\textrm{t}} +\cos\varphi_{\textrm{t}} \cos \varrho_{\textrm{t}}, \sin\vartheta_{\textrm{t}} \sin\varrho_{\textrm{t}} \right ]^{T}, \label{eq3}
\end{align}
where $\varrho_{\textrm{t}}\in \left [ 0,2\pi \right )$ represents the spin angle of the transmitting FIM, which characterizes the angle of the transmitting FIM rotating around its normal direction $\mathbf{k}_{\textrm{t}}$.

Let $M=M_{\textrm{x}}M_{\textrm{y}}$ represent the total number of transmitting antennas, with $M_{\textrm{x}}$ and $M_{\textrm{y}}$ denoting the number of antenna elements along $\mathbf{i}_{\textrm{t}}$ and $\mathbf{j}_{\textrm{t}}$ directions, respectively. Let $\mathbf{q}_{m}\in \mathbb{R}^{3},\ m=1,2,\ldots ,M$ represent the position of the $m$-th transmitting element. Taking the position of the first element, $\mathbf{q}_{1}$, as a reference point, we have
\begin{align}
 \mathbf{q}_{m} = \mathbf{q}_{1}+ x_{m}^{\textrm{t}}\mathbf{i}_{\textrm{t}}+y_{m}^{\textrm{t}}\mathbf{j}_{\textrm{t}},\quad m = 1,2,\ldots, M,
\end{align}
where $x_{m}^{\textrm{t}} = d_{\textrm{t},\textrm{x}}\times \textrm{mod}\left ( m-1,M_{\textrm{x}} \right )$ and $y_{m}^{\textrm{t}} = d_{\textrm{t}, \textrm{y}}\times \left \lfloor \left ( m-1 \right )/M_{\textrm{x}} \right \rfloor$ represent the distance between the $m$-th element and the first element along $\mathbf{i}_{\textrm{t}}$ and $\mathbf{j}_{\textrm{t}}$ directions, respectively. Furthermore, $d_{\textrm{t},\textrm{x}}$ and $d_{\textrm{t},\textrm{y}}$ denote the spacing between adjacent antenna elements in $\mathbf{i}_{\textrm{t}}$ and $\mathbf{j}_{\textrm{t}}$ directions, respectively.

Furthermore, each radiating element on the transmitting FIM can be independently adjusted along its normal direction $\mathbf{k}_{\textrm{t}}$, yielding
\begin{align}
 \tilde{\mathbf{q}}_{m}=\mathbf{q}_{m}+\zeta_{m}\mathbf{k}_{\textrm{t}}, \quad m = 1, 2, \ldots, M,
\end{align}
where $\zeta_{m}$ represents the deformation distance of the $m$-th element\footnote{According to \cite{Nature_2022_Bai_A}, the FIM surface shape can be morphed with a fine resolution of $6$ $\mu$m. In a typical millimeter-wave communication system, this level of adjustment is essentially equivalent to continuous morphing.}, satisfying $-\tilde{\zeta} \leq \zeta_{m} \leq \tilde{\zeta}$, with $\tilde{\zeta}\geq 0$ characterizing the maximum range allowed by the unilateral reversible deformation of the transmitting FIM \cite{Nature_2022_Bai_A}. As a result, the surface shape of the transmitting FIM at the source is characterized by
\begin{align}
 \boldsymbol{\zeta} =\left [ \zeta_{1}, \zeta_{2},\ldots , \zeta_{M} \right ]^{T}\in \mathbb{R}^{M}.
\end{align}

\subsubsection{Receiving FIM}
Similarly, let $\mathbf{i}_{\textrm{r}}\in \mathbb{R}^{3}$ and $\mathbf{j}_{\textrm{r}}\in \mathbb{R}^{3}$ represent the directions of the two sides of the receiving FIM, with $\mathbf{k}_{\textrm{r}}\in \mathbb{R}^{3}$ representing its normal direction. Additionally, let $\varphi_{\textrm{r}}\in \left [ 0,\pi \right )$ and $\vartheta_{\textrm{r}}\in \left [ 0,\pi \right )$ represent the azimuth and elevation angles, respectively, of the normal direction of the receiving FIM. Therefore, the normal vector $\mathbf{k}_{\textrm{r}}$ is expressed as
\begin{align}\label{eq7}
 \mathbf{k}_{\textrm{r}} =& \left [ \sin\vartheta_{\textrm{r}} \cos\varphi_{\textrm{r}} , \sin\vartheta_{\textrm{r}} \sin\varphi_{\textrm{r}} , \cos\vartheta_{\textrm{r}} \right ]^{T}.
\end{align}

Following the same concept as \eqref{eq2} and \eqref{eq3}, $\mathbf{i}_{\textrm{r}}$ and $\mathbf{j}_{\textrm{r}}$ are given by
\begin{align}
 \mathbf{i}_{\textrm{r}} =& \left [ \cos\vartheta_{\textrm{r}} \cos\varphi_{\textrm{r}} \cos\varrho_{\textrm{r}} -\sin\varphi_{\textrm{r}} \sin \varrho_{\textrm{r}}, \cos\vartheta_{\textrm{r}}\right. \notag\\
 & \left. \times \sin\varphi_{\textrm{r}} \cos\varrho_{\textrm{r}} +\cos\varphi_{\textrm{r}} \sin \varrho_{\textrm{r}}, -\sin\vartheta_{\textrm{r}} \cos\varrho_{\textrm{r}} \right ]^{T},\label{eq8}\\
 \mathbf{j}_{\textrm{r}} =& \left [ -\cos\vartheta_{\textrm{r}} \cos\varphi_{\textrm{r}} \sin\varrho_{\textrm{r}} -\sin\varphi_{\textrm{r}} \cos\varrho_{\textrm{r}},-\cos\vartheta_{\textrm{r}}\right. \notag\\
 & \left. \times \sin\varphi_{\textrm{r}} \sin\varrho_{\textrm{r}} +\cos\varphi_{\textrm{r}} \cos \varrho_{\textrm{r}}, \sin\vartheta_{\textrm{r}} \sin\varrho_{\textrm{r}} \right ]^{T},\label{eq9}
\end{align}
where $\varrho_{\textrm{r}}\in \left [ 0,2\pi \right )$ represents the spin angle of the receiving FIM, which characterizes the rotation angle of the receiving FIM around its normal direction $\mathbf{k}_{\textrm{r}}$.

Let $N=N_{\textrm{x}}N_{\textrm{y}}$ represent the total number of receiving antennas, with $N_{\textrm{x}}$ and $N_{\textrm{y}}$ denoting the number of antenna elements along $\mathbf{i}_{\textrm{r}}$ and $\mathbf{j}_{\textrm{r}}$ directions, respectively. Denote the position of the $n$-th receiving element as $\mathbf{p}_{n}\in \mathbb{R}^{3},\ n=1,2,\ldots ,N$ and take the position of the first element, $\mathbf{p}_{1}$, as a reference point. Therefore, we have
\begin{align}
 \mathbf{p}_{n} = \mathbf{p}_{1}+ x_{n}^{\textrm{r}}\mathbf{i}_{\textrm{r}}+y_{n}^{\textrm{r}}\mathbf{j}_{\textrm{r}},\quad n = 1, 2, \ldots, N,
\end{align}
where $x_{n}^{\textrm{r}} = d_{\textrm{r},\textrm{x}}\times \textrm{mod}\left ( n-1,N_{\textrm{x}} \right )$ and $y_{n}^{\textrm{r}} = d_{\textrm{r},\textrm{y}}\times \left \lfloor \left ( n-1 \right )/N_{\textrm{x}} \right \rfloor$ represent the distance between the $n$-th element and the first element along $\mathbf{i}_{\textrm{r}}$ and $\mathbf{j}_{\textrm{r}}$ directions, respectively. The variables $d_{\textrm{r},\textrm{x}}$ and $d_{\textrm{r},\textrm{y}}$ denote the spacings between adjacent antenna elements in $\mathbf{i}_{\textrm{r}}$ and $\mathbf{j}_{\textrm{r}}$ directions, respectively.

Similarly, each radiating element on the receiving FIM can be independently adjusted along its normal direction $\mathbf{k}_{\textrm{r}}$, yielding
\begin{align}
 \tilde{\mathbf{p}}_{n}=\mathbf{p}_{n}+\xi_{n}\mathbf{k}_{\textrm{r}},\quad n=1,2,\ldots,N,
\end{align}
where $\xi_{n}$ represents the deformation distance of the $n$-th element, satisfying $-\tilde{\xi} \leq \xi_{n} \leq \tilde{\xi}$, with $\tilde{\xi}$ denoting the maximum range allowed by the unilateral reversible deformation of the receiving FIM. As a result, the surface shape of the receiving FIM at the destination is characterized by
\begin{align}
 \boldsymbol{\xi} =\left [ \xi_{1}, \xi_{2},\ldots , \xi_{N} \right ]^{T}\in \mathbb{R}^{N}.
\end{align}

\begin{table*}[!t]
\scriptsize
\centering
\caption{Array orientations and rotations of FIMs under eight typical setups.}
\label{tab3}
\renewcommand\arraystretch{1.5}
\tabcolsep=0.2cm
\begin{tabular}{c|c|c|c}
\hline
 $\varphi _{\textrm{t}}=0,\ \vartheta _{\textrm{t}}=0,\ \varrho _{\textrm{t}}=0$ & $\varphi _{\textrm{t}}=0,\ \vartheta _{\textrm{t}}=\pi/2,\ \varrho _{\textrm{t}}=0$ & $\varphi _{\textrm{t}}=\pi/2,\ \vartheta _{\textrm{t}}=0,\ \varrho _{\textrm{t}}=0$ & $\varphi _{\textrm{t}}=\pi/2,\ \vartheta _{\textrm{t}}=\pi/2,\ \varrho _{\textrm{t}}=0$ \\ \hline
\raisebox{-0.9\height}{\includegraphics[width=4cm]{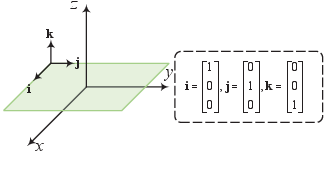}} & \raisebox{-0.9\height}{\includegraphics[width=4cm]{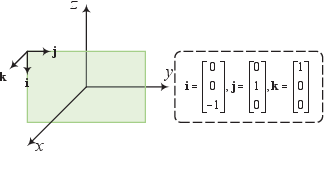}} & \raisebox{-0.9\height}{\includegraphics[width=4cm]{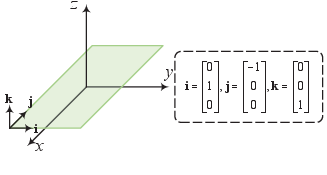}} & \raisebox{-0.9\height}{\includegraphics[width=4cm]{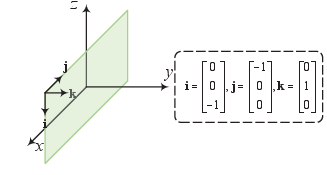}} \\ \hline\hline
$\varphi _{\textrm{t}}=0,\ \vartheta _{\textrm{t}}=0,\ \varrho _{\textrm{t}}=\pi/2$ & $\varphi _{\textrm{t}}=0,\ \vartheta _{\textrm{t}}=\pi/2,\ \varrho _{\textrm{t}}=\pi/2$ & $\varphi _{\textrm{t}}=\pi/2,\ \vartheta _{\textrm{t}}=0,\ \varrho _{\textrm{t}}=\pi/2$ & $\varphi _{\textrm{t}}=\pi/2,\ \vartheta _{\textrm{t}}=\pi/2,\ \varrho _{\textrm{t}}=\pi/2$ \\ \hline
\raisebox{-0.9\height}{\includegraphics[width=4cm]{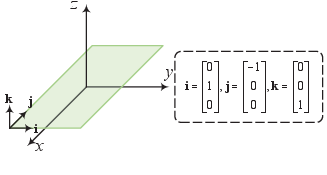}} & \raisebox{-0.9\height}{\includegraphics[width=4cm]{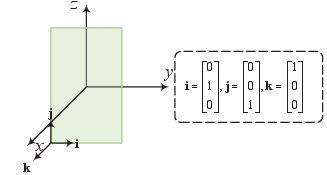}} & \raisebox{-0.9\height}{\includegraphics[width=4cm]{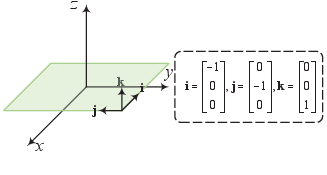}} & \raisebox{-0.9\height}{\includegraphics[width=4cm]{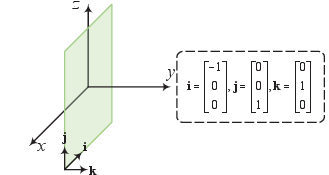}} \\ \hline
\end{tabular}\vspace{-0.5cm}
\end{table*}
\begin{remark}
For illustration, Table \ref{tab3} shows the array orientations and rotations of the transmitting and receiving FIMs under eight typical setups. The corresponding side and normal vectors $\left \{ \mathbf{i}, \mathbf{j}, \mathbf{k} \right \}$ as well as the angle parameters $\left \{ \vartheta, \varphi, \varrho \right \}$ are also shown. For brevity, the subscripts `$\textrm{t}$' and `$\textrm{r}$' indicating transmitting and receiving FIMs have been omitted since the results are applicable for both. It is demonstrated that the normal direction characterized by $\left \{ \vartheta, \varphi \right \}$ and the spin angle characterized by $\varrho$ provide sufficient degrees of freedom (DoF) for characterizing FIMs with arbitrary array orientation and spin angles.
\end{remark}
\begin{remark}
Note that the surface-shape morphing along the normal direction of the transmitting and receiving FIMs can be achieved using the electromagnetic actuation technology presented in \cite{Nature_2022_Bai_A}. Utilizing more complex geometric shapes for the FIMs could further enhance MIMO channel capacity. However, since the transmitting (receiving) FIM remains its array structure, with its two sides aligned with the $\mathbf{i}_{\textrm{t}} \left ( \mathbf{i}_{\textrm{r}}\right )$ and $\mathbf{j}_{\textrm{t}} \left ( \mathbf{j}_{\textrm{r}}\right )$ directions, each antenna element can be viewed as equivalent to the others that move along these directions. Consequently, the benefits of shape morphing in the $\mathbf{i}_{\textrm{t}} \left ( \mathbf{i}_{\textrm{r}}\right )$ and $\mathbf{j}_{\textrm{t}} \left ( \mathbf{j}_{\textrm{r}}\right )$ directions may diminish as the element deployment becomes denser.
\end{remark}

\subsection{Channel Model}
In this paper, we adopt the multipath propagation model of \cite{JSTSP_2016_Heath_An} to characterize the wireless channels. Moreover, we consider a quasi-static block fading environment and focus on a particular block where the channel environment remains approximately constant. Additionally, we consider narrowband transmission over frequency-flat channels.

Note that in contrast to conventional MIMO communication systems, the channels in FIM-aided MIMO systems depend on the 3D surface shapes of the transmitting and receiving FIMs. Specifically, let $\mathbf{H}\left ( \boldsymbol{\zeta}, \boldsymbol{\xi} \right ) \in \mathbb{C}^{N \times M}$ denote the MIMO channel matrix spanning from the transmitting FIM to the receiving FIM. Let $L$ represent the number of scattering clusters, each having $G$ propagation paths. We also assume that all scatterers are located in the far field of the transmitting and receiving FIMs.

Let $\varphi _{l,g}^{\textrm{t}}\in \left [ 0,\pi \right )$ and $\vartheta _{l,g}^{\textrm{t}}\in \left [ 0,\pi \right )$ represent the azimuth and elevation angles of departure (AoD), respectively, for the $g$-th propagation path in the $l$-th scattering cluster \emph{w.r.t.} the transmitting FIM. The corresponding propagation direction is given by
\begin{align}\label{eq13}
 \mathbf{o}_{l,g}^{\textrm{t}} = \left [ \sin\vartheta_{l,g}^{\textrm{t}} \cos\varphi_{l,g}^{\textrm{t}} , \sin\vartheta_{l,g}^{\textrm{t}} \sin\varphi_{l,g}^{\textrm{t}} , \cos\vartheta_{l,g}^{\textrm{t}} \right ] ^{T},
\end{align}
for $g = 1, 2, \ldots, G,\ l = 1, 2, \ldots, L$. Within the $l$-th cluster, the $G$ azimuth and elevation AoDs, $\varphi_{l,g}^{\textrm{t}}$ and $\vartheta_{l,g}^{\textrm{t}}$, are assumed to be uniformly distributed around mean cluster angles of $\varphi_{l}^{\textrm{t}}$ and $\vartheta_{l}^{\textrm{t}}$ respectively. Additionally, they exhibit constant angular spreads -- defined as the standard deviation of the AoAs/AoDs associated with multiple propagation paths within each cluster \cite{TWC_2014_Ayach_Spatially} -- of $\sigma_{\varphi}^{\textrm{t}}$ and $\sigma_{\vartheta}^{\textrm{t}}$, respectively.

According to \eqref{eq13}, the array steering vector $\mathbf{a}_{\textrm{t}} \left ( \varphi_{l,g}^{\textrm{t}} ,\vartheta_{l,g}^{\textrm{t}} \right ) \in \mathbb{C}^{M}$ of the \emph{unmorphed} transmitting FIM is given by
\begin{align}\label{eq14}
 \mathbf{a}_{\textrm{t}}\left ( \varphi_{l,g}^{\textrm{t}}, \vartheta_{l,g}^{\textrm{t}} \right )=&\left [ 1,\ldots , e^{j\kappa \left ( x_{m}^{\textrm{t}}\left \langle \mathbf{i}_{\textrm{t}},\mathbf{o}_{l,g}^{\textrm{t}}\right \rangle+y_{m}^{\textrm{t}}\left \langle \mathbf{j}_{\textrm{t}},\mathbf{o}_{l,g}^{\textrm{t}} \right \rangle \right )}, \right.\notag\\
&\left. \ldots ,e^{j\kappa \left ( x_{M}^{\textrm{t}}\left \langle \mathbf{i}_{\textrm{t}},\mathbf{o}_{l,g}^{\textrm{t}}\right \rangle+y_{M}^{\textrm{t}}\left \langle \mathbf{j}_{\textrm{t}},\mathbf{o}_{l,g}^{\textrm{t}} \right \rangle \right )} \right ]^{T},
\end{align}
where $\kappa = 2\pi / \lambda $ represents the wavenumber, with $\lambda $ denoting the radio wavelength.

Furthermore, the transmitting FIM is capable of imposing additional 3D deformation to adjust the position of each antenna deposited on it. Specifically, the additional multiplicative response component $\mathbf{f}_{\textrm{t}} \left ( \boldsymbol{\zeta}, \varphi_{l,g}^{\textrm{t}} ,\vartheta_{l,g}^{\textrm{t}} \right ) \in \mathbb{C}^{M}$ is given by
\begin{align}\label{eq15}
 \mathbf{f}_{\textrm{t}}\left ( \boldsymbol{\zeta}, \varphi_{l,g}^{\textrm{t}}, \vartheta_{l,g}^{\textrm{t}} \right )=\left [ 1,\ldots , e^{j\kappa \zeta_{m}\left \langle \mathbf{k}_{\textrm{t}},\mathbf{o}_{l,g}^{\textrm{t}} \right \rangle }, \ldots ,e^{j\kappa \zeta_{M}\left \langle \mathbf{k}_{\textrm{t}},\mathbf{o}_{l,g}^{\textrm{t}} \right \rangle } \right ]^{T}.
\end{align}

As a result, the array steering vector $\tilde{\mathbf{a}}_{\textrm{t}}\left ( \boldsymbol{\zeta}, \varphi_{l,g}^{\textrm{t}}, \vartheta_{l,g}^{\textrm{t}} \right )\in \mathbb{C}^{M}$ of the transmitting FIM can be represented as
\begin{align}\label{eq16}
 \tilde{\mathbf{a}}_{\textrm{t}}\left ( \boldsymbol{\zeta}, \varphi_{l,g}^{\textrm{t}}, \vartheta_{l,g}^{\textrm{t}} \right )=\mathbf{a}_{\textrm{t}}\left ( \varphi_{l,g}^{\textrm{t}}, \vartheta_{l,g}^{\textrm{t}} \right )\odot\mathbf{f}_{\textrm{t}}\left ( \boldsymbol{\zeta}, \varphi_{l,g}^{\textrm{t}}, \vartheta_{l,g}^{\textrm{t}} \right ),
\end{align}
for $g=1,2,\ldots,G$ and $l=1,2,\ldots,L$.

Note that when considering $\varphi _{\textrm{t}}=0,\ \vartheta _{\textrm{t}}=0,\ \varrho _{\textrm{t}}=0$, we have $\mathbf{i}_{\textrm{t}}=\left [ 1,0,0 \right ]^{T},\ \mathbf{j}_{\textrm{t}}=\left [ 0,1,0 \right ]^{T},\ \mathbf{k}_{\textrm{t}}=\left [ 0,0,1 \right ]^{T}$. In this case, \eqref{eq14} and \eqref{eq15} simplify accordingly to \eqref{eq17_0} and
\begin{figure*}[!t]
\begin{align}\label{eq17_0}
\mathbf{a}_{\textrm{t}}\left ( \varphi_{l,g}^{\textrm{t}}, \vartheta_{l,g}^{\textrm{t}} \right )=\left [ 1,\ldots , e^{j\kappa \left ( x_{m}^{\textrm{t}}\sin\vartheta_{l,g}^{\textrm{t}} \cos\varphi_{l,g}^{\textrm{t}}+y_{m}^{\textrm{t}}\sin\vartheta_{l,g}^{\textrm{t}} \sin\varphi_{l,g}^{\textrm{t}} \right )}, \ldots ,e^{j\kappa \left ( x_{M}^{\textrm{t}}\sin\vartheta_{l,g}^{\textrm{t}} \cos\varphi_{l,g}^{\textrm{t}}+y_{M}^{\textrm{t}}\sin\vartheta_{l,g}^{\textrm{t}} \sin\varphi_{l,g}^{\textrm{t}} \right )} \right ]^{T}.
\end{align}
\end{figure*}
\begin{align}
 \mathbf{f}_{\textrm{t}}\left ( \boldsymbol{\zeta}, \varphi_{l,g}^{\textrm{t}}, \vartheta_{l,g}^{\textrm{t}} \right )=\left [ 1,\ldots , e^{j\kappa \zeta_{m}\cos\vartheta_{l,g}^{\textrm{t}} }, \ldots ,e^{j\kappa \zeta_{M}\cos\vartheta_{l,g}^{\textrm{t}} } \right ]^{T}.
\end{align}

Similarly, let $\varphi _{l,g}^{\textrm{r}}\in \left [ 0,\pi \right )$ and $\vartheta _{l,g}^{\textrm{r}}\in \left [ 0,\pi \right )$ represent the azimuth and elevation angles of arrival (AoA), respectively, for the $g$-th propagation path in the $l$-th scattering cluster \emph{w.r.t.} the receiving FIM. The corresponding propagation direction $\mathbf{o}_{l,g}^{\textrm{r}}\in \mathbb{R}^{N}$ is expressed as
\begin{align}\label{eq17}
 \mathbf{o}_{l,g}^{\textrm{r}} = \left [ \sin\vartheta_{l,g}^{\textrm{r}} \cos\varphi_{l,g}^{\textrm{r}} , \sin\vartheta_{l,g}^{\textrm{r}} \sin\varphi_{l,g}^{\textrm{r}} , \cos\vartheta_{l,g}^{\textrm{r}} \right ] ^{T},
\end{align}
for $g = 1, 2, \ldots, G,\ l = 1, 2, \ldots, L$. Additionally, the $G$ azimuth and elevation AoAs, $\varphi_{l,g}^{\textrm{r}}$ and $\vartheta_{l,g}^{\textrm{r}}$, within the $l$-th cluster are uniformly distributed around the mean cluster angles of $\varphi_{l}^{\textrm{r}}$ and $\vartheta_{l}^{\textrm{r}}$, with constant angular spreads of $\sigma_{\varphi}^{\textrm{r}}$ and $\sigma_{\vartheta}^{\textrm{r}}$, respectively.

According to \eqref{eq17}, the array steering vector $\mathbf{a}_{\textrm{r}}\left ( \varphi_{l,g}^{\textrm{r}}, \vartheta_{l,g}^{\textrm{r}} \right )\in \mathbb{C}^{N}$ of the \emph{unmorphed} receiving FIM is given by
\begin{align}\label{eq18}
 \mathbf{a}_{\textrm{r}}\left ( \varphi_{l,g}^{\textrm{r}} ,\vartheta_{l,g}^{\textrm{r}} \right )=\left [ 1,\ldots , e^{j\kappa \left ( x_{n}^{\textrm{r}}\left \langle \mathbf{i}_{\textrm{r}},\mathbf{o}_{l,g}^{\textrm{r}} \right \rangle+y_{n}^{\textrm{r}}\left \langle \mathbf{j}_{\textrm{r}},\mathbf{o}_{l,g}^{\textrm{r}} \right \rangle \right )}, \right.\notag\\
\left. \ldots ,e^{j\kappa \left ( x_{N}^{\textrm{r}}\left \langle \mathbf{i}_{\textrm{r}},\mathbf{o}_{l,g}^{\textrm{r}} \right \rangle +y_{N}^{\textrm{r}}\left \langle \mathbf{j}_{\textrm{r}},\mathbf{o}_{l,g}^{\textrm{r}} \right \rangle \right )} \right ]^{T}.
\end{align}

Furthermore, let $\mathbf{f}_{\textrm{r}} \left ( \boldsymbol{\xi}, \varphi_{l,g}^{\textrm{r}} ,\vartheta_{l,g}^{\textrm{r}} \right ) \in \mathbb{C}^{N}$ represent the additional response component imposed by adjusting the position of each antenna deposited on the receiving FIM, which can be explicitly expressed as
\begin{align}\label{eq19}
 \mathbf{f}_r\left ( \boldsymbol{\xi}, \varphi_{l,g}^{\textrm{r}}, \vartheta_{l,g}^{\textrm{r}} \right ) = \left [ 1,\ldots , e^{j\kappa \xi_{n}\left \langle \mathbf{k}_{\textrm{r}},\mathbf{o}_{l,g}^{\textrm{r}} \right \rangle }, \ldots ,e^{j\kappa \xi_{N}\left \langle \mathbf{k}_{\textrm{r}},\mathbf{o}_{l,g}^{\textrm{r}} \right \rangle } \right ]^{T}. 
\end{align}

As a result, the array steering vector $\tilde{\mathbf{a}}_{\textrm{r}}\left ( \boldsymbol{\xi}, \varphi_{l,g}^{\textrm{r}}, \vartheta_{l,g}^{\textrm{r}} \right )\in \mathbb{C}^{N}$ of the receiving FIM can be represented as
\begin{align}\label{eq20}
\tilde{\mathbf{a}}_{\textrm{r}}\left ( \boldsymbol{\xi}, \varphi_{l,g}^{\textrm{r}}, \vartheta_{l,g}^{\textrm{r}} \right )=\mathbf{a}_{\textrm{r}}\left ( \varphi_{l,g}^{\textrm{r}}, \vartheta_{l,g}^{\textrm{r}} \right )\odot\mathbf{f}_{\textrm{r}}\left ( \boldsymbol{\xi}, \varphi_{l,g}^{\textrm{r}}, \vartheta_{l,g}^{\textrm{r}} \right ),
\end{align}
for $g=1,2,\ldots,G$ and $l=1,2,\ldots,L$.

Note that when considering $\varphi _{\textrm{r}}=0,\ \vartheta _{\textrm{r}}=0,\ \varrho _{\textrm{r}}=0$, we have $\mathbf{i}_{\textrm{r}}=\left [ 1,0,0 \right ]^{T},\ \mathbf{j}_{\textrm{r}}=\left [ 0,1,0 \right ]^{T},\ \mathbf{k}_{\textrm{r}}=\left [ 0,0,1 \right ]^{T}$. In this case, \eqref{eq18} and \eqref{eq19} simplify accordingly to \eqref{eq23_0} and
\begin{figure*}[!t]
\begin{align}\label{eq23_0}
 \mathbf{a}_{\textrm{r}}\left ( \varphi_{l,g}^{\textrm{r}}, \vartheta_{l,g}^{\textrm{r}} \right )=\left [ 1,\ldots , e^{j\kappa \left ( x_{n}^{\textrm{r}}\sin\vartheta_{l,g}^{\textrm{r}} \cos\varphi_{l,g}^{\textrm{r}}+y_{n}^{\textrm{r}}\sin\vartheta_{l,g}^{\textrm{r}} \sin\varphi_{l,g}^{\textrm{r}} \right )}, \ldots ,e^{j\kappa \left ( x_{N}^{\textrm{r}}\sin\vartheta_{l,g}^{\textrm{r}} \cos\varphi_{l,g}^{\textrm{r}}+y_{N}^{\textrm{r}}\sin\vartheta_{l,g}^{\textrm{r}} \sin\varphi_{l,g}^{\textrm{r}} \right )} \right ]^{T}.
\end{align}
\end{figure*}
\begin{align}
 \mathbf{f}_{\textrm{r}}\left ( \boldsymbol{\zeta}, \varphi_{l,g}^{\textrm{r}}, \vartheta_{l,g}^{\textrm{r}} \right )=\left [ 1,\ldots , e^{j\kappa \zeta_{n}\cos\vartheta_{l,g}^{\textrm{r}} }, \ldots ,e^{j\kappa \zeta_{N}\cos\vartheta_{l,g}^{\textrm{r}} } \right ]^{T}.
\end{align}

Upon combining \eqref{eq16} and \eqref{eq20}, the narrowband MIMO channel $\mathbf{H}\left ( \boldsymbol{\zeta},\boldsymbol{\xi} \right )$ from the transmitting FIM to the receiving FIM can be written as
\begin{align}\label{eq21}
 \mathbf{H}\left ( \boldsymbol{\zeta},\boldsymbol{\xi} \right )=\sum_{l=1}^{L}\sum_{g=1}^{G}\varsigma _{l,g}\tilde{\mathbf{a}}_{\textrm{r}}\left ( \boldsymbol{\xi}, \varphi_{l,g}^{\textrm{r}}, \vartheta_{l,g}^{\textrm{r}} \right )\tilde{\mathbf{a}}_{\textrm{t}}^{H}\left ( \boldsymbol{\zeta}, \varphi_{l,g}^{\textrm{t}}, \vartheta_{l,g}^{\textrm{t}} \right ),
\end{align}
where $\varsigma _{l,g} \in \mathbb{C},\ g=1,2,\ldots,G,\ l=1,2,\ldots,L$ represents the complex gain of the $g$-th propagation path in the $l$-th scattering cluster. Specifically, $\varsigma _{l,g}$ is modeled as an independent and identically distributed (i.i.d.) CSCG variable, satisfying $\varsigma _{l, g}\sim \mathcal{CN}\left ( 0,\rho _{l}^{2}/G \right )$, where $\rho _{l}^{2}$ denotes the average power of the $l$-th cluster \cite{TWC_2014_Ayach_Spatially}. Furthermore, let $\beta^{2}$ characterize the path loss between the source and the destination. Consequently, we have $\sum\nolimits_{l=1}^{L}\rho _{l}^{2}=\beta^{2}$.

For the sake of brevity, we define $\mathbf{A}_{\textrm{t}}\in\mathbb{C}^{M\times LG}$, $\mathbf{F}_{\textrm{t}}\left ( \boldsymbol{\zeta} \right )\in\mathbb{C}^{M\times LG}$, $\mathbf{A}_{\textrm{r}}\in\mathbb{C}^{N\times LG}$, $\mathbf{F}_{\textrm{r}}\left ( \boldsymbol{\xi} \right )\in\mathbb{C}^{N\times LG}$, and $\boldsymbol{\varsigma}\in\mathbb{C}^{LG\times LG}$ as
\begin{align}
 \mathbf{A}_{\textrm{t}}=&\left [ \mathbf{a}_{\textrm{t}}\left ( \varphi_{1,1}^{\textrm{t}}, \vartheta_{1,1}^{\textrm{t}} \right ),\ldots ,\mathbf{a}_{\textrm{t}}\left ( \varphi_{1,G}^{\textrm{t}}, \vartheta_{1,G}^{\textrm{t}} \right ),\ldots ,\right. \notag\\
 &\left.\mathbf{a}_{\textrm{t}}\left ( \varphi_{L,1}^{\textrm{t}}, \vartheta_{L,1}^{\textrm{t}} \right ),\ldots ,\mathbf{a}_{\textrm{t}}\left ( \varphi_{L,G}^{\textrm{t}}, \vartheta_{L,G}^{\textrm{t}} \right ) \right ],\\
 \mathbf{F}_{\textrm{t}}\left ( \boldsymbol{\zeta} \right )=&\left [ \mathbf{f}_{\textrm{t}}\left ( \boldsymbol{\zeta}, \varphi_{1,1}^{\textrm{t}}, \vartheta_{1,1}^{\textrm{t}} \right ),\ldots ,\mathbf{f}_{\textrm{t}}\left ( \boldsymbol{\zeta}, \varphi_{1,G}^{\textrm{t}}, \vartheta_{1,G}^{\textrm{t}} \right ),\ldots ,\right. \notag\\
 &\left.\mathbf{f}_{\textrm{t}}\left ( \boldsymbol{\zeta}, \varphi_{L,1}^{\textrm{t}}, \vartheta_{L,1}^{\textrm{t}} \right ),\ldots ,\mathbf{f}_{\textrm{t}}\left ( \boldsymbol{\zeta}, \varphi_{L,G}^{\textrm{t}}, \vartheta_{L,G}^{\textrm{t}} \right ) \right ],\\
 \mathbf{A}_{\textrm{r}}=&\left [ \mathbf{a}_r\left ( \varphi_{1,1}^{\textrm{r}}, \vartheta_{1,1}^{\textrm{r}} \right ),\ldots ,\mathbf{a}_r\left ( \varphi_{1,G}^{\textrm{r}}, \vartheta_{1,G}^{\textrm{r}} \right ),\ldots ,\right. \notag\\
 &\left.\mathbf{a}_r\left ( \varphi_{L,1}^{\textrm{r}}, \vartheta_{L,1}^{\textrm{r}} \right ),\ldots ,\mathbf{a}_r\left ( \varphi_{L,G}^{\textrm{r}}, \vartheta_{L,G}^{\textrm{r}} \right ) \right ],\\
 \mathbf{F}_{\textrm{r}}\left ( \boldsymbol{\xi} \right )=&\left [ \mathbf{f}_r\left ( \boldsymbol{\xi}, \varphi_{1,1}^{\textrm{r}}, \vartheta_{1,1}^{\textrm{r}} \right ),\ldots ,\mathbf{f}_r\left ( \boldsymbol{\xi}, \varphi_{1,G}^{\textrm{r}}, \vartheta_{1,G}^{\textrm{r}} \right ),\ldots ,\right. \notag\\
 &\left.\mathbf{f}_r\left ( \boldsymbol{\xi}, \varphi_{L,1}^{\textrm{r}}, \vartheta_{L,1}^{\textrm{r}} \right ),\ldots ,\mathbf{f}_r\left ( \boldsymbol{\xi}, \varphi_{L,G}^{\textrm{r}}, \vartheta_{L,G}^{\textrm{r}} \right ) \right ],\\
 \boldsymbol{\varsigma }=&\textrm{diag}\left ( \varsigma_{1,1},\ldots ,\varsigma_{1,G},\ldots ,\varsigma_{L,1},\ldots ,\varsigma_{L,G} \right ).
\end{align}

Hence, the matrix form of the FIM-aided MIMO channel in \eqref{eq21} is given as
\begin{align}\label{eq25}
 \mathbf{H}\left ( \boldsymbol{\zeta},\boldsymbol{\xi} \right )=\left [ \mathbf{A}_{\textrm{r}}\odot \mathbf{F}_{\textrm{r}}\left ( \boldsymbol{\xi} \right ) \right ]\boldsymbol{\varsigma } \left [ \mathbf{A}_{\textrm{t}}\odot \mathbf{F}_{\textrm{t}}\left ( \boldsymbol{\zeta} \right ) \right ]^{H}.
\end{align}

\subsection{Signal Model}
Let $\mathbf{t}\in \mathbb{C}^{M }$ and $\mathbf{T} \triangleq \mathbb{E}\left ( \mathbf{t}\mathbf{t}^{H} \right ) \in \mathbb{C}^{M\times M}$ denote the transmit signal vector and the transmit signal covariance matrix, respectively, satisfying $\mathbf{T} \succeq \mathbf{0}$. We consider an average sum power constraint at the source given by $\mathbb{E} \left ( \left \| \mathbf{t} \right \|^{2} \right ) = \textrm{tr} \left ( \mathbf{T} \right )\leq P_{\textrm{t}}$. After passing through the wireless channel, the complex baseband signal vector $\mathbf{r}\in \mathbb{C}^{N }$ received at the destination is given by
\begin{align}
 \mathbf{r}=\mathbf{H}\left ( \boldsymbol{\zeta},\boldsymbol{\xi} \right )\mathbf{t}+\mathbf{v},
\end{align}
where $\mathbf{v} \sim \mathcal{CN}\left ( \mathbf{0},\sigma^{2} \mathbf{I}_{N} \right )$ represents the CSCG noise vector at the destination, with $\sigma^{2}$ denoting the average noise power.

Hence, the MIMO channel capacity is given by
\begin{align}\label{eq29}
 \underset{\mathbf{T},\, \boldsymbol{\zeta},\, \boldsymbol{\xi}}{\max}\underbrace{\log_{2}\det\left ( \mathbf{I}_{N}+\frac{1}{\sigma ^{2}}\mathbf{H}\left ( \boldsymbol{\zeta},\boldsymbol{\xi} \right )\mathbf{T}\mathbf{H}^{H}\left ( \boldsymbol{\zeta},\boldsymbol{\xi} \right ) \right )}_{C},
\end{align}
in bits per second per Hertz (bps/Hz). In conventional MIMO communication systems with RAAs ($\boldsymbol{\zeta}=\mathbf{0}$ and $\boldsymbol{\xi}=\mathbf{0}$), the capacity is fully determined by the channel matrix $\mathbf{H}\left ( \mathbf{0},\mathbf{0} \right )$. By contrast, for MIMO communications between a pair of FIMs, the channel capacity $\mathbf{H}\left ( \boldsymbol{\zeta},\boldsymbol{\xi} \right )$ also depends on the 3D surface-shape configurations of the transmitting FIM $\boldsymbol{\zeta}$ and receiving FIM $\boldsymbol{\xi}$, which provide extra design DoFs for further improving the channel capacity in a more energy-efficient way.
\subsection{Problem Formulation}
In this paper, we aim to maximize the capacity of a MIMO channel between a pair of FIMs by jointly optimizing three factors: \emph{i)} the 3D surface shape $\boldsymbol{\zeta}$ of the transmitting FIM, \emph{ii)} the 3D surface shape $\boldsymbol{\xi}$ of the receiving FIM, and \emph{iii)} the transmit covariance matrix $\mathbf{T}$, subject to the morphing range of these two FIMs and a total power constraint at the source. Specifically, the joint optimization problem is formulated as
\begin{subequations}\label{eq30}
\begin{alignat}{2}
&\max_{\mathbf{T},\, \boldsymbol{\zeta},\, \boldsymbol{\xi}} &\quad& \log_{2}\det\left ( \mathbf{I}_{N}+\frac{1}{\sigma ^{2}}\mathbf{H}\left ( \boldsymbol{\zeta},\boldsymbol{\xi} \right )\mathbf{T}\mathbf{H}^{H}\left ( \boldsymbol{\zeta},\boldsymbol{\xi} \right ) \right ) \label{eq30a}\\
&\textrm{s.t.} & & \textrm{tr}\left ( \mathbf{T} \right )\leq P_{\textrm{t}},\quad \mathbf{T}\succeq \mathbf{0}, \label{eq30c}\\
& & & \boldsymbol{\zeta} =\left [ \zeta_{1}, \zeta_{2},\ldots , \zeta_{M} \right ]^{T}, \label{eq30d}\\
& & & \boldsymbol{\xi} =\left [ \xi_{1}, \xi_{2},\ldots , \xi_{N} \right ]^{T}, \label{eq30e}\\
& & & -\tilde{\zeta} \leq \zeta_{m}\leq \tilde{\zeta},\quad m=1,2,\ldots,M, \label{eq30f}\\
& & & -\tilde{\xi}\leq \xi_{n}\leq \tilde{\xi},\quad n=1,2,\ldots,N, \label{eq30g}
\end{alignat}
\end{subequations}
where \eqref{eq30c} characterizes the transmit power constraint, while \eqref{eq30d} -- \eqref{eq30g} characterize the constraints on adjusting the deformation range for each antenna on the transmitting and receiving FIMs.

Note that problem \eqref{eq30} is challenging to solve optimally due to the fact that the objective function in \eqref{eq30a} is non-concave \emph{w.r.t.} the surface-shape configurations $\boldsymbol{\zeta}$ and $\boldsymbol{\xi}$ of the transmitting and receiving FIMs. Moreover, $\boldsymbol{\zeta}$ and $\boldsymbol{\xi}$ are complicatedly interacted with the transmit covariance matrix $\mathbf{T}$ in the objective function \eqref{eq30a}, making it impossible to solve for them independently. In the next section, we will decompose problem \eqref{eq30} into two subproblems and solve them efficiently.

\begin{remark}
To reveal the theoretical capacity limit of FIM-aided MIMO communications, we assume perfect knowledge of all channel parameters, i.e., $\varphi_{l,g}^{\textrm{r}}$, $\vartheta_{l,g}^{\textrm{r}}$, $\varphi_{l,g}^{\textrm{t}}$, $\vartheta_{l,g}^{\textrm{t}}$, and $\varsigma _{l,g}$ at both the source and destination. These channel parameters can be probed by utilizing existing parametric channel estimation methods, such as the compressed sensing technique and the space-alternating generalized expectation maximization algorithm described in \cite{TWC_2018_Shahmansoori_Position, TWC_2023_An_Fundamental}. Moreover, the surface shapes of the transmitting and receiving FIMs can be potentially optimized to attain higher channel estimation accuracy. This topic is beyond the scope of this paper and is left to our future research.
\end{remark}

\begin{remark}
Note that the 3D surface shapes of the transmitting and receiving FIMs are only updated on the timescale of the channel's coherence block, which is much longer than the symbol duration. This dynamic flexibility can be readily achieved using advanced electromagnetic actuation technology \cite{Nature_2022_Bai_A}. Specifically, the positions of the antennas on the flexible substrate can be adjusted by reconfiguring the distributed Lorentz forces from electrical currents passing through a static magnetic field. This enables the FIM to morph its surface shape precisely and promptly as required.
\end{remark}

\section{The Proposed Block Coordinate Descent Algorithm}\label{sec3}
In this section, we propose an efficient BCD algorithm to sub-optimally solve the problem formulated in \eqref{eq30}. Specifically, we solve two subproblems for optimizing the transmit covariance matrix $\mathbf{T}$ or the surface shapes of the transmitting and receiving FIMs $\left \{ \boldsymbol{\zeta},\ \boldsymbol{\xi} \right \}$ respectively, with the other set of variables being fixed. We solve these two subproblems iteratively until convergence is reached.

\subsection{Transmit Covariance Optimization with Given FIMs' Surface-Shape Configurations $\left \{ \hat{\boldsymbol{\zeta}},\ \hat{\boldsymbol{\xi}}\right \}$}\label{sec3_1}
In this subproblem, we aim to optimize the transmit covariance matrix $\mathbf{T}$ given a tentative surface-shape configuration $\left \{ \hat{\boldsymbol{\zeta}},\ \hat{\boldsymbol{\xi}}\right \}$ of the transmitting and receiving FIMs. In this case, the MIMO channel $\mathbf{H}\left ( \hat{\boldsymbol{\zeta}},\ \hat{\boldsymbol{\xi}} \right )$ from the source to the destination is determined by \eqref{eq25}, and the original problem in \eqref{eq30} is reduced to optimizing the transmit covariance matrix as in a conventional MIMO system, with the optimal $\mathbf{T}$ being determined by the eigenmode transmission \cite{BOOK_2005_Tse_Fundamentals}.

Specifically, let $\mathbf{H}^{H}\left ( \hat{\boldsymbol{\zeta}},\ \hat{\boldsymbol{\xi}} \right )\mathbf{H}\left ( \hat{\boldsymbol{\zeta}},\ \hat{\boldsymbol{\xi}} \right )=\mathbf{U}\mathbf{\Lambda} \mathbf{U}^{H}$ denote the eigenvalue decomposition of $\mathbf{H}^{H}\left ( \hat{\boldsymbol{\zeta}},\ \hat{\boldsymbol{\xi}} \right )\mathbf{H}\left ( \hat{\boldsymbol{\zeta}},\ \hat{\boldsymbol{\xi}} \right )$, with $\boldsymbol{\Lambda }=\textrm{diag}\left ( \lambda _{1}^2,\lambda _{2}^2,\ldots ,\lambda _{M}^2 \right )\in \mathbb{R}^{M \times M}$ representing a diagonal matrix containing the corresponding eigenvalues. The optimal transmit covariance matrix $\mathbf{T}^{o}$ is thus given by\footnote{In practical MIMO systems, $\mathbf{U}\left ( \mathbf{P}^{o}\right )^{1/2}$ is utilized as the transmit precoding matrix, which means that the superimposed signal of multiple symbols is transmitted from each antenna element.}
\begin{align}\label{eq31}
 \mathbf{T}^{o}=\mathbf{U}\mathbf{P}^{o}\mathbf{U}^{H},
\end{align}
where we have $\mathbf{P}^{o}=\textrm{diag}\left ( p_{1}^{o},p_{2}^{o},\ldots ,p_{M}^{o} \right )\in \mathbb{R}^{M\times M}$, and $p_m^{o}$ represents the amount of power allocated to the $m$-th data stream. Specifically, the optimal value of $p_m^{o}$ can be obtained using the water-filling strategy, which gives:
\begin{align}\label{eq32}
 p_{m}^{o}=\max\left ( \mu - \sigma ^{2}/\lambda _{m}^{2},0 \right ),\quad m = 1,2,\ldots ,M,
\end{align}
where $\mu$ is a threshold such that $\sum\nolimits_{m=1}^{M}p_{m}^{o}=P_{\textrm{t}}$, which can be obtained by applying the bisection search \cite{TCOM_2022_An_Low}.

\subsection{Surface-Shape Morphing with Given Transmit Covariance Matrix $\hat{\mathbf{T}}$}\label{sec3_2}
Next, we aim to obtain the optimal surface-shape configuration $\left \{ \boldsymbol{\zeta},\ \boldsymbol{\xi} \right \}$ in \eqref{eq30} with given transmit covariance matrix $\hat{\mathbf{T}}$. Specifically, substituting $\hat{\mathbf{T}}$ into \eqref{eq30} yields
\begin{subequations}\label{eq33}
\begin{alignat}{2}
&\max_{\boldsymbol{\zeta},\ \boldsymbol{\xi}} &\quad& \log_{2}\det\left ( \mathbf{I}_{N}+\frac{1}{\sigma ^{2}}\mathbf{H}\left ( \boldsymbol{\zeta},\boldsymbol{\xi} \right )\hat{\mathbf{T}}\mathbf{H}^{H}\left ( \boldsymbol{\zeta},\boldsymbol{\xi} \right ) \right ) \label{eq33a}\\
&\textrm{s.t.} & & \eqref{eq30d},\ \eqref{eq30e},\ \eqref{eq30f},\ \eqref{eq30g}. \label{eq33b}
\end{alignat}
\end{subequations}

Since an optimal solution to \eqref{eq33} is still difficult to obtain, instead we leverage the gradient ascent algorithm to find a sub-optimal solution. Specifically, given the surface shapes of the transmitting and receiving FIMs found from the previous iteration as the initial point, we adjust their surface-shape configurations towards the direction of the gradient to gradually increase the channel capacity. The gradient ascent algorithm consists of two major steps: \emph{i)} gradient calculation, and \emph{ii)} surface-shape update.
\subsubsection{\bf Gradient Calculation}
The gradient ascent method requires calculating the gradients of the objective function $C$ \emph{w.r.t.} the surface-shape configurations $\boldsymbol{\zeta}$ and $\boldsymbol{\xi}$ of the transmitting and receiving FIMs, which can be obtained by using \emph{Propositions 1 and 2}, respectively.

\begin{proposition}
 The gradient of $C$ \emph{w.r.t.} $\boldsymbol{\xi}$, denoted by $\nabla_{\boldsymbol{\xi} } C$, is given by
\begin{align}
 \nabla_{\boldsymbol{\xi} } C = -\frac{2}{ \ln 2}\textrm{Diag}\left [ \mathbf{B}_{\textrm{r}}^{-1}\odot\Im \left ( \mathbf{S}_{\textrm{r}}\right ) \right ], \label{eq34}
\end{align}
where the matrices $\mathbf{S}_{\textrm{r}}\in\mathbb{C}^{N \times N}$, $\mathbf{B}_{\textrm{r}}\in\mathbb{C}^{N \times N}$, $\mathbf{O}_{\textrm{t}}\in\mathbb{C}^{LG \times LG}$, and $\mathbf{K}_{\textrm{r}}\in\mathbb{R}^{LG \times LG}$ are defined by
\begin{align}
 \mathbf{S}_{\textrm{r}}=&\left [ \mathbf{A}_{\textrm{r}}\odot \mathbf{F}_{\textrm{r}}\left ( \boldsymbol{\xi} \right ) \right ]\mathbf{K}_{\textrm{r}}\mathbf{O}_{\textrm{t}}\left [ \mathbf{A}_{\textrm{r}}\odot \mathbf{F}_{\textrm{r}}\left ( \boldsymbol{\xi} \right ) \right ]^{H}, \label{eq35}\\
 \mathbf{B}_{\textrm{r}}=&\mathbf{I}_{N}+\left [ \mathbf{A}_{\textrm{r}}\odot \mathbf{F}_{\textrm{r}}\left ( \boldsymbol{\xi} \right ) \right ]\mathbf{O}_{\textrm{t}}\left [ \mathbf{A}_{\textrm{r}}\odot \mathbf{F}_{\textrm{r}}\left ( \boldsymbol{\xi} \right )\right ]^{H} , \label{eq36}\\
 \mathbf{O}_{\textrm{t}}=&\frac{1}{\sigma ^{2}}\boldsymbol{\varsigma } \left [ \mathbf{A}_{\textrm{t}}\odot \mathbf{F}_{\textrm{t}}\left ( \boldsymbol{\zeta} \right ) \right ]^{H}\hat{\mathbf{T}}\left [ \mathbf{A}_{\textrm{t}}\odot \mathbf{F}_{\textrm{t}}\left ( \boldsymbol{\zeta} \right ) \right ]\boldsymbol{\varsigma }^{H},\\
 \mathbf{K}_{\textrm{r}}=&\textrm{diag}\left ( \kappa \left \langle \mathbf{k}_{\textrm{r}},\mathbf{o}_{1,1}^{\textrm{r}} \right \rangle ,\ldots,\kappa \left \langle \mathbf{k}_{\textrm{r}},\mathbf{o}_{1,G}^{\textrm{r}} \right \rangle ,\ldots ,\right . \notag\\
 &\left. \kappa \left \langle \mathbf{k}_{\textrm{r}},\mathbf{o}_{L,1}^{\textrm{r}} \right \rangle ,\ldots, \kappa \left \langle \mathbf{k}_{\textrm{r}},\mathbf{o}_{L,G}^{\textrm{r}} \right \rangle \right ),
\end{align}
respectively.
\end{proposition}

\emph{Proof:} Please refer to Appendix \ref{A1}. $\hfill\blacksquare$

\begin{proposition}
The gradient of $C$ \emph{w.r.t.} $\boldsymbol{\zeta}$, denoted by $\nabla_{\boldsymbol{\zeta} } C$, is given by
\begin{align}
 \nabla_{\boldsymbol{\zeta} } C = -\frac{2}{ \ln 2}\textrm{Diag}\left [ \mathbf{B}_{\textrm{t}}^{-1}\odot \Im \left ( \mathbf{S}_{\textrm{t}}\right ) \right ].\label{eq39}
\end{align}
where matrices $\mathbf{S}_{\textrm{t}}\in\mathbb{C}^{M \times M}$, $\mathbf{B}_{\textrm{t}}\in\mathbb{C}^{M \times M}$, $\mathbf{O}_{\textrm{r}}\in\mathbb{C}^{LG \times LG}$, and $\mathbf{K}_{\textrm{t}}\in\mathbb{R}^{LG \times LG}$ are defined by
\begin{align}
 \mathbf{S}_{\textrm{t}}=&\hat{\mathbf{T}}\left [ \mathbf{A}_{\textrm{t}}\odot \mathbf{F}_{\textrm{t}}\left ( \boldsymbol{\zeta} \right ) \right ]\mathbf{K}_{\textrm{t}}\mathbf{O}_{\textrm{r}}\left [ \mathbf{A}_{\textrm{t}}\odot \mathbf{F}_{\textrm{t}}\left ( \boldsymbol{\zeta} \right ) \right ]^{H}, \label{eq40}\\
 \mathbf{B}_{\textrm{t}}=&\mathbf{I}_{M}+\hat{\mathbf{T}}\left [ \mathbf{A}_{\textrm{t}}\odot \mathbf{F}_{\textrm{t}}\left ( \boldsymbol{\zeta} \right ) \right ]\mathbf{O}_{\textrm{r}}\left [ \mathbf{A}_{\textrm{t}}\odot \mathbf{F}_{\textrm{t}}\left ( \boldsymbol{\zeta} \right )\right ]^{H} ,\label{eq41}\\
 \mathbf{O}_{\textrm{r}}=&\frac{1}{\sigma ^{2}}\boldsymbol{\varsigma }^{H} \left [ \mathbf{A}_{\textrm{r}}\odot \mathbf{F}_{\textrm{r}}\left ( \boldsymbol{\xi} \right ) \right ]^{H}\left [ \mathbf{A}_{\textrm{r}}\odot \mathbf{F}_{\textrm{r}}\left ( \boldsymbol{\xi} \right ) \right ]\boldsymbol{\varsigma },\\
 \mathbf{K}_{\textrm{t}}=&\textrm{diag}\left ( \kappa \left \langle \mathbf{k}_{\textrm{t}},\mathbf{o}_{1,1}^{\textrm{t}} \right \rangle ,\ldots,\kappa \left \langle \mathbf{k}_{\textrm{t}},\mathbf{o}_{1,G}^{\textrm{t}} \right \rangle ,\ldots ,\right . \notag\\
 &\left. \kappa \left \langle \mathbf{k}_{\textrm{t}},\mathbf{o}_{L,1}^{\textrm{t}} \right \rangle ,\ldots, \kappa \left \langle \mathbf{k}_{\textrm{t}},\mathbf{o}_{L,G}^{\textrm{t}} \right \rangle \right ),
\end{align}
respectively.
\end{proposition}

\emph{Proof:} Please refer to Appendix \ref{A2}. $\hfill\blacksquare$

\subsubsection{\bf Surface-Shape Morphing}
At each iteration, the 3D surface shapes of the transmitting and receiving FIMs are updated by
\begin{align}
 \boldsymbol{\zeta}&\leftarrow \boldsymbol{\zeta} + \epsilon \nabla_{\boldsymbol{\zeta}} C,\label{eq42}\\
 \boldsymbol{\xi}&\leftarrow \boldsymbol{\xi} + \epsilon \nabla_{\boldsymbol{\xi}} C,\label{eq43}
\end{align}
where $\epsilon > 0$ represents the step size, which is determined by applying a backtracking line search to prevent the overshooting effect \cite{JSAC_2023_An_Stacked}.

Additionally, a projection process is imposed on the positions obtained from \eqref{eq42} and \eqref{eq43} for scaling them into the allowable morphing ranges of the FIMs, i.e.,
\begin{align}
 \zeta_{m}&=\max\left ( \min\left ( \zeta_{m},\tilde{\zeta} \right ),-\tilde{\zeta} \right ),\quad m= 1,2,\ldots, M,\\
 \xi_{n}&=\max\left ( \min\left ( \xi_{n},\tilde{\xi} \right ),-\tilde{\xi} \right ),\quad n= 1,2,\ldots, N.
\end{align}

By repeatedly updating the surface shapes several times according to \eqref{eq42} and \eqref{eq43}, the optimal surface shapes $\boldsymbol{\zeta}^{\textrm{o}}$ and $\boldsymbol{\xi}^{\textrm{o}}$ of the transmitting and receiving FIMs can be obtained.

\begin{algorithm}[!t]
\caption{The Proposed Block Coordinate Descent Algorithm.}
\begin{algorithmic}[1]
\STATE \textbf{Input}: $\varsigma _{l,g},\ \varphi_{l,g}^{\textrm{t}},\ \vartheta_{l,g}^{\textrm{t}},\ \varphi_{l,g}^{\textrm{r}},\ \vartheta_{l,g}^{\textrm{r}},\ g=1,2,\ldots,G,\ l=1,2,\ldots,L$, $\tilde{\zeta},\ \tilde{\xi}$, $P_{\textrm{t}},\ \sigma^2$.
\STATE Initialize the surface shapes $\boldsymbol{\zeta}$ and $\boldsymbol{\xi}$ of the transmitting and receiving FIMs.
\STATE \textbf{Repeat}
\STATE \hspace{0.5cm} Optimize the transmit covariance matrix for the given\\
\hspace{0.5cm} $\hat{\boldsymbol{\zeta}}$ and $\hat{\boldsymbol{\xi}}$ using the eigenmode transmission method\\
\hspace{0.5cm} presented in Section \ref{sec3_1}.
\STATE \hspace{0.5cm} Optimize the surface shapes of the transmitting and\\
\hspace{0.5cm} receiving FIMs for the given $\hat{\mathbf{T}}$ using the gradient\\
\hspace{0.5cm} ascent method presented in Section \ref{sec3_2}.\\
\STATE \textbf{Until} The fractional increase of the MIMO channel capacity falls below a preset threshold or the maximum tolerable number of iterations is reached.
\STATE \textbf{Output}: $\boldsymbol{\xi}^{\textrm{o}},\ \boldsymbol{\zeta}^{\textrm{o}},\ \mathbf{T}^{\textrm{o}}$.
\end{algorithmic}
\label{alg1}
\end{algorithm}
In a nutshell, the proposed BCD algorithm solves the original problem by alternately performing water-filling power allocation and gradient-based surface-shape morphing in an iterative manner, until the objective function in \eqref{eq30a} achieves convergence. The overall procedure of the BCD algorithm is outlined in Algorithm \ref{alg1}. The proposed BCD algorithm is guaranteed to converge to at least a locally optimal solution for two reasons:
\begin{itemize}
 \item \emph{Firstly}, the value of the objective function in \eqref{eq30a} is non-decreasing as the iteration proceeds by selecting an appropriate step size according to the backtracking line search.
 \item \emph{Secondly}, the objective function is upper bounded due to the power constraint in \eqref{eq30c}.
\end{itemize}

\begin{remark}
When dealing with frequency-selective channels, it is essential for the surface shapes of the transmitting and receiving FIMs to accommodate the requirements of multiple sub-carriers. Specifically, the array steering vectors represented in \eqref{eq14}, \eqref{eq15}, \eqref{eq18}, and \eqref{eq19} need to be revised to incorporate the sub-carrier frequency component. Additionally, by including the sub-carrier index in the objective function \eqref{eq30a} and calculating total channel capacity across all sub-carriers, the proposed BCD method can be effectively adapted to address the FIM surface-shape morphing and the transmit covariance optimization in frequency-selective scenarios. Note that the water-filling method described in \eqref{eq32} should be modified to implement a joint spatial-frequency water-filling approach \cite{BOOK_2005_Tse_Fundamentals}. A comprehensive performance evaluation of FIM in frequency-selective scenarios will constitute our future research topic.
\end{remark}

\begin{figure}[!t]
\centering
\subfloat[$\tilde{\xi} = \lambda/2$, $L = 1$.]{\includegraphics[width=4.25cm]{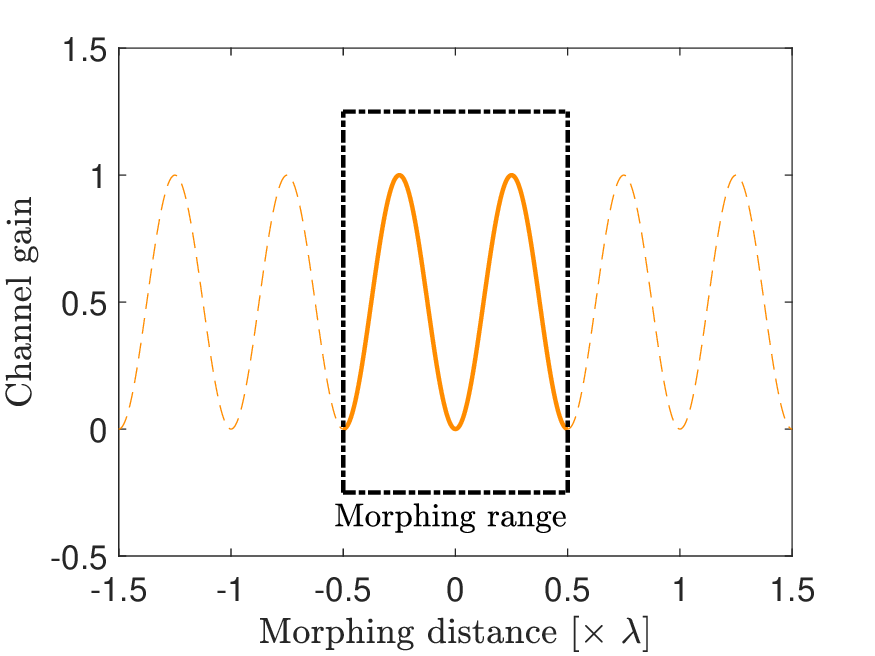}
\label{fig_3_1}}
\subfloat[$\tilde{\xi} = \lambda/2$, $L = 10$.]{\includegraphics[width=4.25cm]{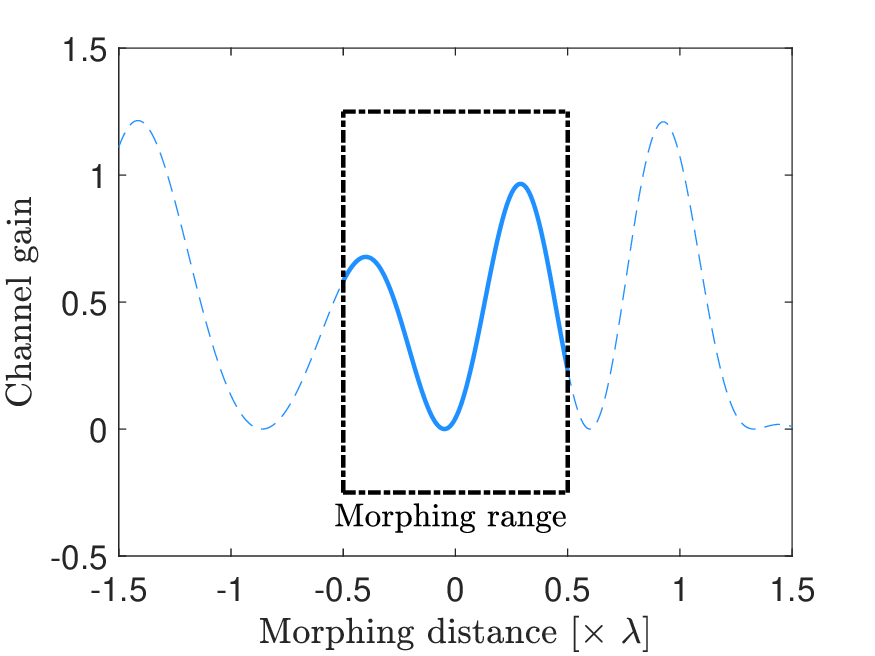}
\label{fig_3_2}}\\
\subfloat[$\tilde{\xi} = \lambda$, $L = 1$.]{\includegraphics[width=4.25cm]{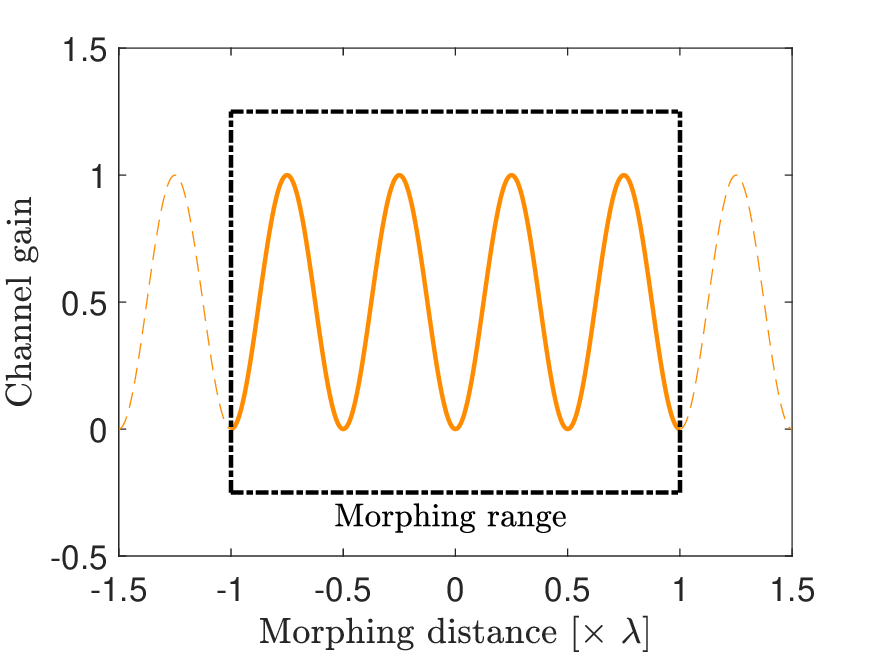}
\label{fig_3_3}}
\subfloat[$\tilde{\xi} = \lambda$, $L = 10$.]{\includegraphics[width=4.25cm]{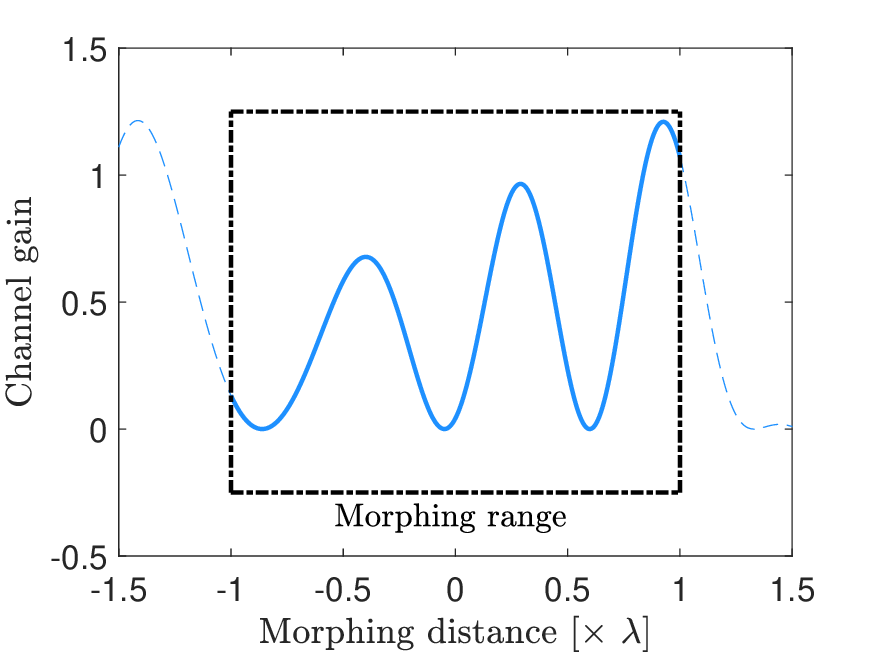}
\label{fig_3_4}}
\caption{Channel gain versus the morphing distance, which demonstrates that the number of peaks increases with the morphing range.}\vspace{-0.5cm}
\label{fig_3}
\end{figure}

\subsection{Initialization of FIM Surface Shapes}
To further enhance the robustness of the BCD algorithm, in our simulations, a heuristic initialization method is utilized by randomly configuring the 3D surface shapes of the transmitting and receiving FIMs multiple times and selecting the configuration that maximizes the MIMO channel capacity. To gain basic insights into the number of random initializations required, we next examine the effective channel gain of a single receiving antenna element versus its morphing distance. For simplicity, we assume a single propagation path in each scattering cluster, as multiple propagation paths within the same cluster have nearly the same periods when projected onto the normal direction of the receiving FIM. Since the received signal is essentially a sum of multiple sine waves with different periods, the number of initializations required is influenced by the number of peaks in the superimposed channel gain observed within a given range. This, in turn, depends on two factors: \emph{i)} the number of scattering clusters $L$, and \emph{ii)} the morphing range $\tilde{\xi}$. With more scatterers, more sine waves are superimposed, while a larger morphing range corresponds to a larger observation range.

Specifically, Fig. \ref{fig_3}(a) illustrates the normalized channel gain when considering a single scatterer with normal incidence, which yields the minimum period of the incident signal gain along the normal direction of the receiving FIM, i.e., $\lambda$ (m). This indicates that the maximum number of peaks in the channel gain is limited to $2$ within the morphing range of $\lambda$. In Fig. \ref{fig_3}(b), we increase the number of scatterers to $L = 10$, each with a random incident angle. Since the periods of sine waves coming from other directions are greater than $\lambda$ (specifically, $\lambda/\left \langle \mathbf{k}_{\textrm{r}},\mathbf{o}_{l,g}^{\textrm{r}} \right \rangle > \lambda$), the number of peaks in the superimposed channel gain remains relatively unaffected by the number of scattering clusters. Assuming a uniform distribution of peaks, the number of peaks within the morphing range $2 \tilde{\xi}$ can be approximately given by $4 \tilde{\xi} /\lambda $ (see Figs. \ref{fig_3}(c) and \ref{fig_3}(d)). Note that this rough estimate of the number of peaks is just for initializing the BCD algorithm. A more accurate analysis of the number of peaks in the superimposed signal is a complex mathematical problem that is beyond the scope of this paper.

\subsection{Complexity Analysis}
Next, we evaluate the complexity of the proposed BCD algorithm. In each iteration, the water-filling power allocation algorithm of Section \ref{sec3_1} has a complexity order of $\mathcal{O}\left (\min\left ( MN^{2}+M^{2}N \right )\right )$. Moreover, the complexity of updating the surface shape in Section \ref{sec3_2} largely depends on calculating the gradients of $\nabla_{\boldsymbol{\xi} } C$ and $\nabla_{\boldsymbol{\zeta} } C$. Note that some matrix multiplications are repeated in calculating gradients, which only have to be calculated once. Specifically, the complexities of calculating $\left [ \mathbf{A}_{\textrm{r}}\odot \mathbf{F}_{\textrm{r}}\left ( \boldsymbol{\xi} \right ) \right ]\boldsymbol{\varsigma } \left [ \mathbf{A}_{\textrm{t}}\odot \mathbf{F}_{\textrm{t}}\left ( \boldsymbol{\zeta} \right ) \right ]^{H}$ and $\hat{\mathbf{T}}\mathbf{H}^{H}\left ( \boldsymbol{\zeta},\boldsymbol{\xi} \right )$ are $LG\left ( MN+M+N \right )$ and $M^2N$, respectively. With these intermediate variables, the complexity of calculating $\mathbf{S}_{\textrm{r}}$ in \eqref{eq35}, $\mathbf{B}_{\textrm{r}}$ in \eqref{eq36}, and $\nabla_{\boldsymbol{\xi} } C$ in \eqref{eq34} are $MN^2$, $MN^2$, and $2N^3$, respectively. Similarly, the complexities of calculating $\mathbf{S}_{\textrm{t}}$ in \eqref{eq40}, $\mathbf{B}_{\textrm{t}}$ in \eqref{eq41}, and $\nabla_{\boldsymbol{\zeta} } C$ in \eqref{eq39} are $2M^2N$, $M^2N$, and $2M^3$, respectively.

As a result, the overall complexity order of the BCD algorithm is
\begin{align}
 \mathcal{O}\left ( I\left [\min\left ( MN^{2}+M^{2}N \right )+LG\left ( MN+M+N \right )\right.\right. \notag\\
 \left.\left.+2MN^2+2N^3+4M^2N+2M^3\right ]\right ),
\end{align}
which is polynomial in $M$ and $N$, where $I$ represents the number of outer iterations.

\begin{figure}[!t]
\centering
\includegraphics[width = 7.5cm]{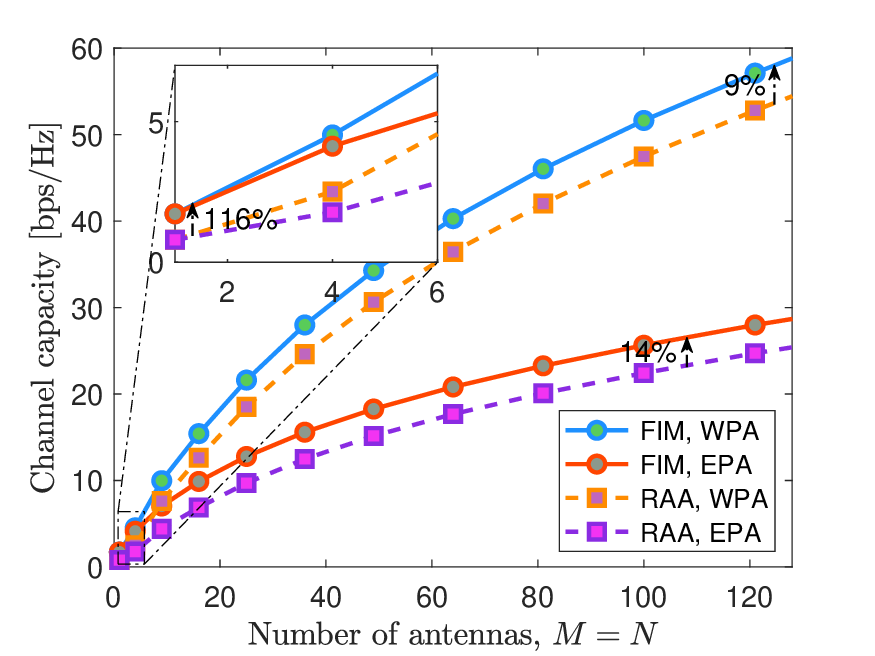}
\caption{Channel capacity $C$ versus the number of antennas $M = N$.}
\label{fig_4}
\end{figure}

\begin{figure}[!t]
\centering
\includegraphics[width = 7.5cm]{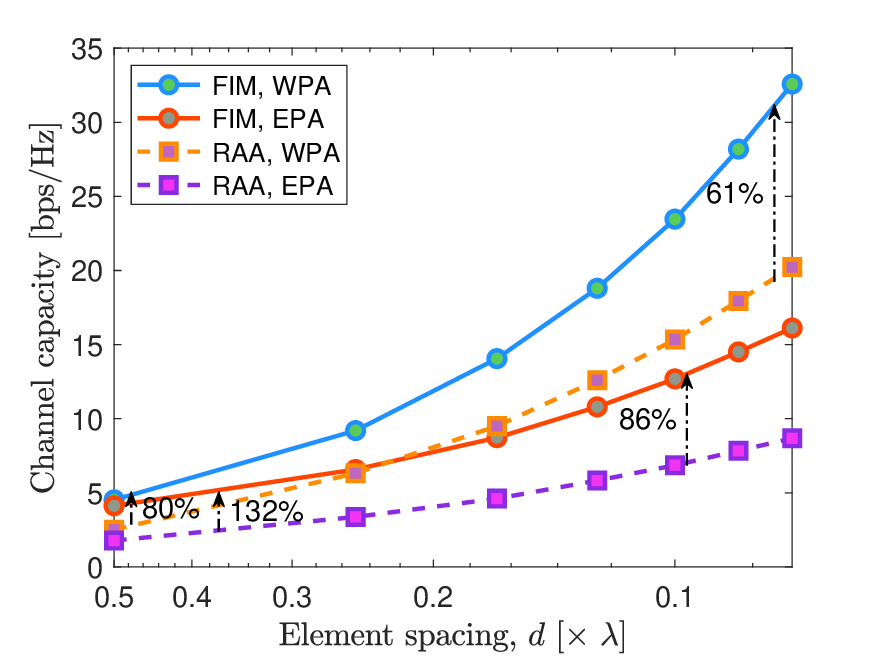}
\caption{Channel capacity $C$ versus the antenna element spacing $d$.}\vspace{-0.5cm}
\label{fig_5}
\end{figure}
\section{Numerical Results}\label{sec4}
\subsection{Simulation Setup}\label{sec4_1} 
In this section, simulation results are presented for examining the performance benefits of utilizing FIMs as MIMO transceivers and the proposed BCD algorithm for maximizing the capacity of FIM-aided MIMO systems. We consider a 3D Cartesian coordinate system, where the locations of the reference antennas at the source and destination are set as $\left ( 0\ \text{m}, 0\ \text{m}, 10\ \text{m} \right )$ and $\left ( 0\ \text{m}, 100\ \text{m}, 0\ \text{m} \right )$, respectively. Both the source and destination are equipped with an FIM, with $M=M_{\textrm{x}}M_{\textrm{y}}$ antennas at the source and $N = N_{\textrm{x}}N_{\textrm{y}}$ antennas at the destination. The antenna spacings at the source and destination are $\left \{ d_{\textrm{t},\textrm{x}},\ d_{\textrm{t},\textrm{y}} \right \}$ and $\left \{ d_{\textrm{r},\textrm{x}},\ d_{\textrm{r},\textrm{y}} \right \}$, respectively. For simplicity, we define $d \triangleq d_{\textrm{t},\textrm{x}} = d_{\textrm{t},\textrm{y}} = d_{\textrm{r},\textrm{x}} = d_{\textrm{r},\textrm{y}} $ in our simulations. Furthermore, the orientations of the transmitting and receiving FIMs are determined by \eqref{eq1} -- \eqref{eq3} and \eqref{eq7} -- \eqref{eq9}, respectively. The FIMs' orientations are configured by setting $\varphi _{\textrm{t}} = \varphi _{\textrm{r}} = \pi/2,\ \vartheta _{\textrm{t}}= \vartheta _{\textrm{r}} =3\pi/4,\ \varrho _{\textrm{t}}=\varrho _{\textrm{r}} = 0$, unless otherwise specified.

\begin{figure}[!t]
\centering
\includegraphics[width = 7.5cm]{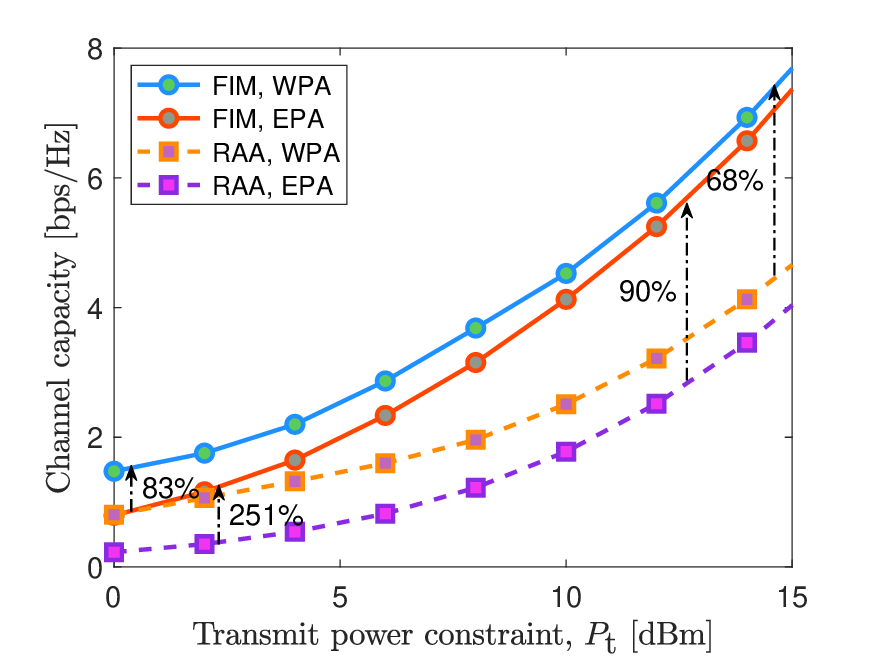}
\caption{Channel capacity $C$ versus the transmit power constraint $P_\textrm{t}$.}
\label{fig_6}
\end{figure}
\begin{figure}[!t]
\centering
\includegraphics[width = 7.5cm]{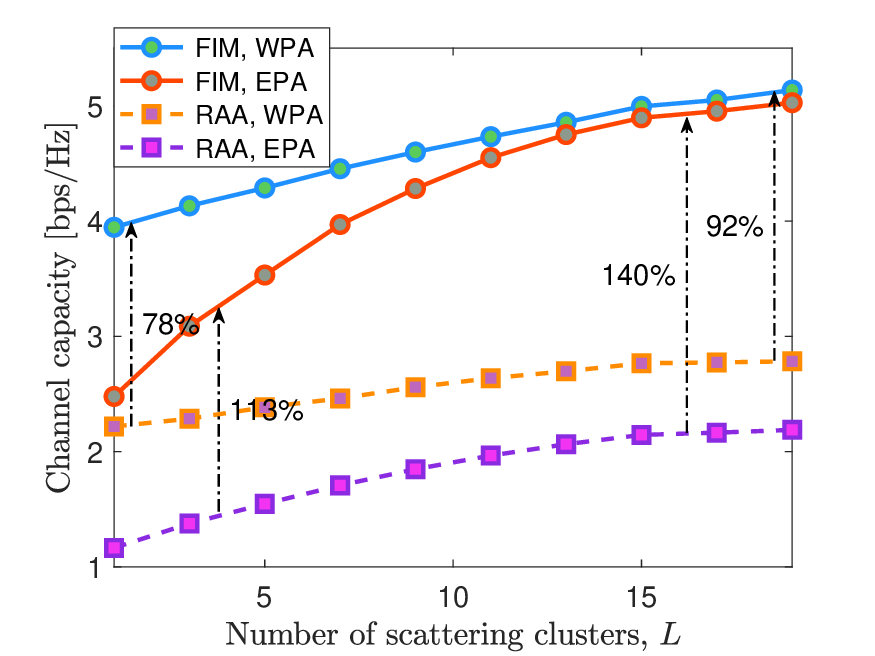}
\caption{Channel capacity $C$ versus the number of scattering clusters $L$.}\vspace{-0.5cm}
\label{fig_7}
\end{figure}
Furthermore, the distance-dependent path loss of the wireless channel is modeled as $\beta ^{2}=\beta _{0}^{2}\left ( d/d_{0} \right )^{-\alpha}$ \cite{TCOM_2015_Rappaport_Wideband}, where $\beta _{0}^2 = -60$ dB denotes the path loss at the reference distance of $d_0 = 1$ m, $\alpha = 2.2$ is the path loss exponent used in our simulations. Additionally, we assume that the $L$ cluster centers are uniformly distributed, satisfying $\varphi _{l}^{\textrm{t}},\ \vartheta_{l}^{\textrm{t}},\ \varphi _{l}^{\textrm{r}},\ \vartheta _{l}^{\textrm{r}}\sim \mathcal{U}\left [ 0,\pi \right )$ for $l=1,2,\ldots,L$. For simplicity, we further assume $\sigma_{\varphi}^{\textrm{t}} = \sigma_{\varphi}^{\textrm{r}} \triangleq \sigma_{\varphi}$ and $\sigma_{\vartheta}^{\textrm{t}} = \sigma_{\vartheta}^{\textrm{r}} \triangleq \sigma_{\vartheta}$. All scatterers experience the same channel gain, yielding $\rho _{l}^{2}=\beta^{2}/L$ for $l=1,2,\ldots,L$. Moreover, the system operates at $28$ GHz with a bandwidth of $100$ MHz. Assuming a noise power spectral density of $-174$ dBm/Hz yields an average noise power of $\sigma^2 = -94$ dBm.

Moreover, four transmission schemes are considered for evaluating the performance of the FIM:
\begin{itemize}
 \item \textbf{FIM, WPA:} A pair of FIMs are utilized as transceivers. The power allocation and surface shapes of the FIMs are alternately updated by using the water-filling strategy and the gradient ascent method, respectively, as presented in Section \ref{sec3}.
 \item \textbf{FIM, EPA:} The equal power allocation regime is utilized at the source. The transmitting and receiving FIM surface shapes are updated by using the gradient ascent method presented in Section \ref{sec3_2}.
 \item \textbf{RAA, WPA \cite{JSAC_2003_Goldsmith_Capacity}:} Conventional MIMO transceivers based on RAAs are employed, while the classic water-filling power allocation is utilized at the source.
 \item \textbf{RAA, EPA \cite{JSAC_2003_Goldsmith_Capacity}:} The equal power allocation is utilized at the conventional RAA-based MIMO transmitter.
\end{itemize}
For each scheme, the transmit covariance matrix is determined by \eqref{eq31}. For the proposed BCD algorithm, the maximum tolerable number of iterations is set to $50$, and the convergence threshold in terms of the fractional increase in channel capacity is set to $-30$ dB. All simulation results are obtained by averaging over $100$ independent channel realizations.

\subsection{Performance versus MIMO System Parameters}\label{sec64}
First, Fig. \ref{fig_4} evaluates the channel capacity versus the number of transmitting and receiving antennas. For simplicity, we consider the same number of antennas at the source and destination, i.e., $M = N$. The antenna spacing is set to $d = \lambda/2$. Moreover, we assume that there are $L = 8$ scattering clusters in the propagation environment, each having $G = 4$ paths. The maximum 3D morphing ranges of the two FIMs are set to $\tilde{\zeta} = \tilde{\xi} = \lambda/2$. The angle spread in each cluster is set to $\sigma_{\varphi} = \sigma_{\vartheta} = \pi/128$. As expected, higher channel capacity is achieved for a larger number of antennas. Additionally, the gap between the two power allocation schemes widens as the number of antennas increases. This is because the MIMO channel may become rank-deficient for a large number of antennas. Compared to equal power allocation, allocating more power to the channels of high quality provides significant gains. Under all setups considered, the FIMs that are capable of morphing their surface shapes consistently outperform the conventional benchmark schemes, with at least a $9$\% capacity improvement being observed.

Furthermore, Fig. \ref{fig_5} examines the FIM-aided MIMO channel capacity versus the antenna element spacing $d$. In this analysis, both the transmitter and receiver have an array area of $\lambda/2 \times \lambda/2$. As the spacing $d$ decreases, a larger number of antenna elements are packed into the designated area, approaching the concept of holographic MIMO communications \cite{CL_2023_An_A1, TWC_2022_Pizzo_Fourier}. Our results, as shown in Fig. \ref{fig_5}, demonstrate that higher channel capacity is achieved as the density of antenna elements increases. Nonetheless, in conventional holographic MIMO communication systems, transmitting and receiving antennas are arranged along the array directions, i.e., $\mathbf{i}_{\textrm{t}}\left ( \mathbf{i}_{\textrm{r}} \right )$ and $\mathbf{j}_{\textrm{t}}\left ( \mathbf{j}_{\textrm{r}} \right )$. In contrast, FIMs have the capability of morphing their surface shapes perpendicular to the antenna array, which leads to further improvements in channel gain. Notably, as the element spacing falls below $0.1\lambda$, leveraging FIMs can yield a capacity improvement of over $60$\% compared to the conventional RAAs.

Next, we evaluate the channel capacity of an FIM-aided point-to-point MIMO system versus the maximum transmit power $P_\textrm{t}$, as shown in Fig. \ref{fig_6}. We consider two square FIMs with four elements and $d = \lambda/2$. All other parameters are the same as in Fig. \ref{fig_3}. As seen from Fig. \ref{fig_6}, by flexibly morphing the surface shapes of the FIMs, the MIMO channel capacity is increased by about $83$\% and $68$\% in the low and high signal-to-noise-ratio (SNR) regions, respectively, compared to their RAA counterparts. Under all SNR setups, the water-filling power allocation outperforms the equal power allocation, while the performance gap gradually narrows as the transmit power increases, since the water-filling threshold is negligible when the power budget is adequate. Interestingly, using FIMs can further narrow the performance gap between these two power allocation schemes. This is due to the fact that the FIMs improve the channel quality and implicitly drive the MIMO communication system to operate in a higher SNR region.

\begin{figure}[!t]
\centering
\subfloat[RAAs.]{\includegraphics[width=4.25cm]{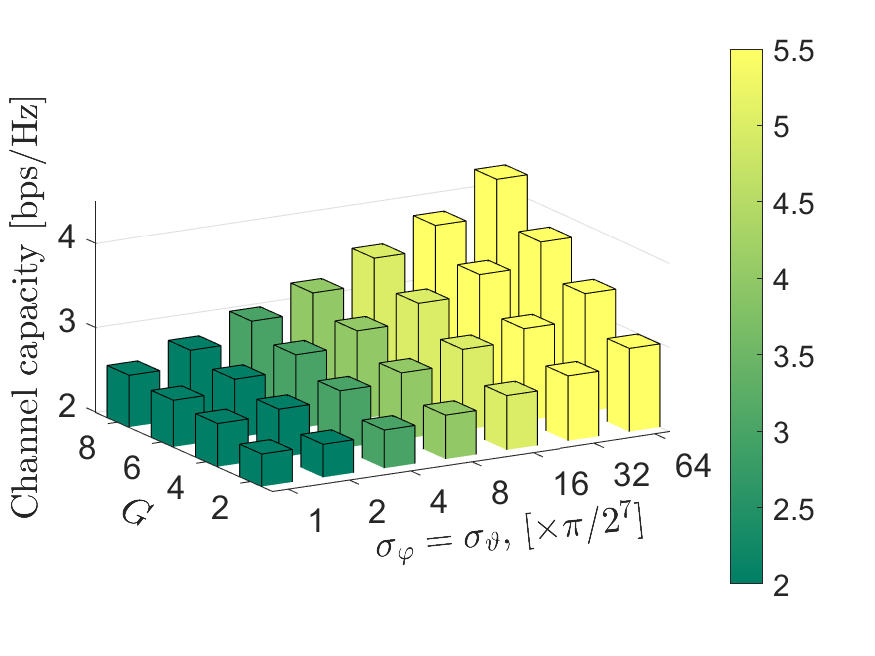}
\label{fig_8_1}}
\subfloat[FIMs.]{\includegraphics[width=4.25cm]{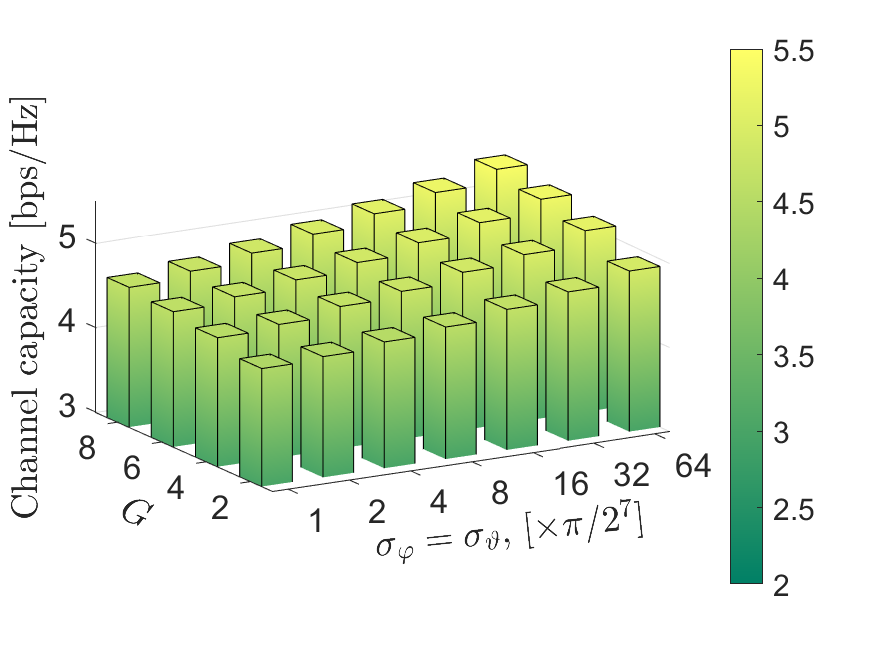}
\label{fig_8_2}}
\caption{Channel capacity versus the number $G$ of propagation paths in each cluster and the angular spread ${\sigma}_{\varphi} = {\sigma}_{\vartheta}$.}
\label{fig_8}
\end{figure}

\begin{figure}[!t]
\centering
\subfloat[Equal power allocation.]{\includegraphics[width=4.25cm]{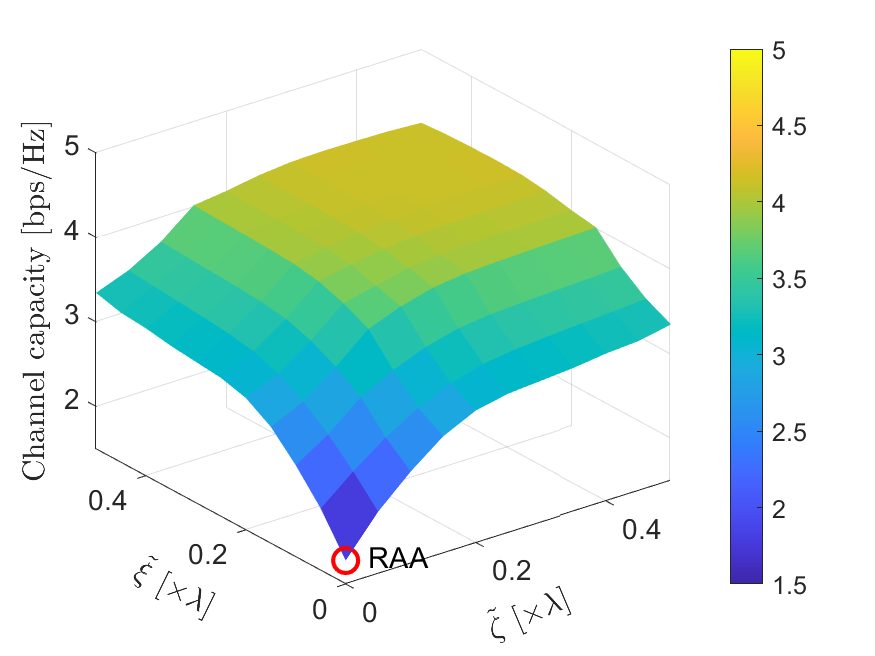}
\label{fig_9_1}}
\subfloat[Water-filling power allocation.]{\includegraphics[width=4.25cm]{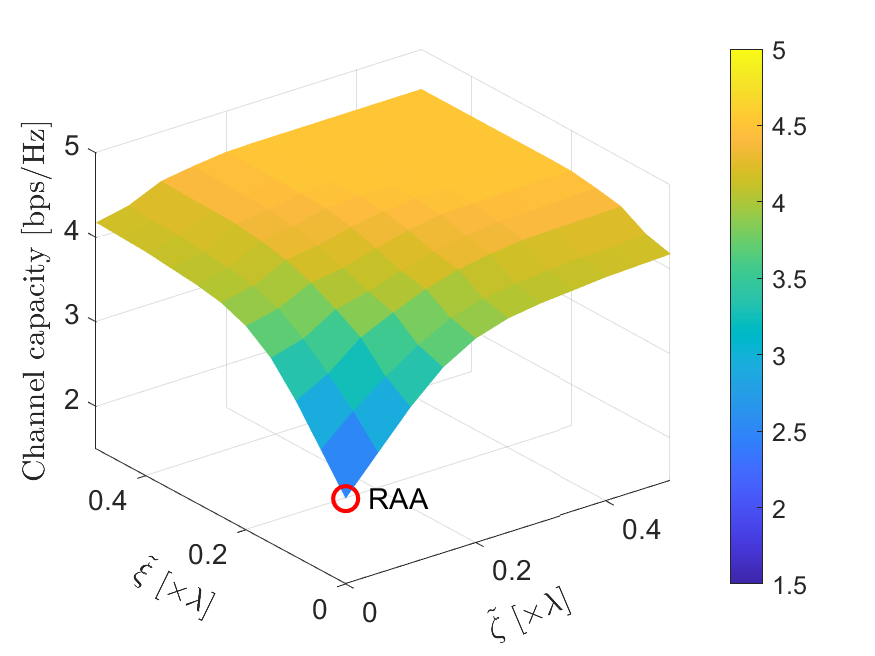}
\label{fig_9_2}}
\caption{Channel capacity $C$ versus the morphing range of the FIMs.}\vspace{-0.5cm}
\label{fig_9}
\end{figure}

Fig. \ref{fig_7} examines the performance as the number of scattering clusters in the environment increases. The downlink transmit power constraint is set to $P_\textrm{t} = 10$ dBm, while all other parameters remain the same as in Fig. \ref{fig_3}. It is demonstrated that as $L$ increases, the MIMO channel capacity improves since the signal components from multiple propagation paths are more likely to form a favorable profile across the array. Moreover, the FIMs outperform conventional RAAs by morphing their surface shapes to find the optimal configurations experiencing shallow fading. Notably, the performance gain offered by the FIMs becomes more substantial, as the number of scattering clusters increases. Specifically, increasing the number of clusters from $L = 1$ to $L = 19$ widens the performance gain from $78$\% to $92$\%, implicitly indicating the importance of multipath propagation for FIM-aided wireless communications.

Fig. \ref{fig_8} assesses the impact of the scattering environment, including the number ($G$) of propagation paths in each scattering cluster and the angular spread (${\sigma}_{\varphi}$, ${\sigma}_{\vartheta}$). Specifically, a larger value of $G$ means that there are more scattering paths in a cluster, while a larger value of ${\sigma}_{\varphi}$ and ${\sigma}_{\vartheta}$ means that the scatterers are distributed over a wider area. For simplicity, we assume ${\sigma}_{\varphi} = {\sigma}_{\vartheta}$. Additionally, we assume that there are $L = 8$ scattering clusters, and all other parameters are the same as in Fig. \ref{fig_3}. Comparing Figs. \ref{fig_8}(a) and \ref{fig_8}(b) shows the channel capacity improvements of using FIMs. Furthermore, increasing both the number of paths and the angular spread would further enhance the channel capacity. Interestingly, for a small value of ${\sigma}_{\varphi}$ and ${\sigma}_{\vartheta}$, increasing $G$ would not result in significant capacity improvement. This is because multiple propagation paths in a small cluster behave more like a single path. By contrast, for a large angle spread, increasing the number of propagation paths would result in the channels interfering constructively and destructively in the surroundings of the FIM more frequently, thus giving the FIMs more opportunity to improve the channel quality by morphing their 3D surface shapes.

\begin{figure}[!t]
\centering
\subfloat[$M = N = 4$.]{\includegraphics[width=4.25cm]{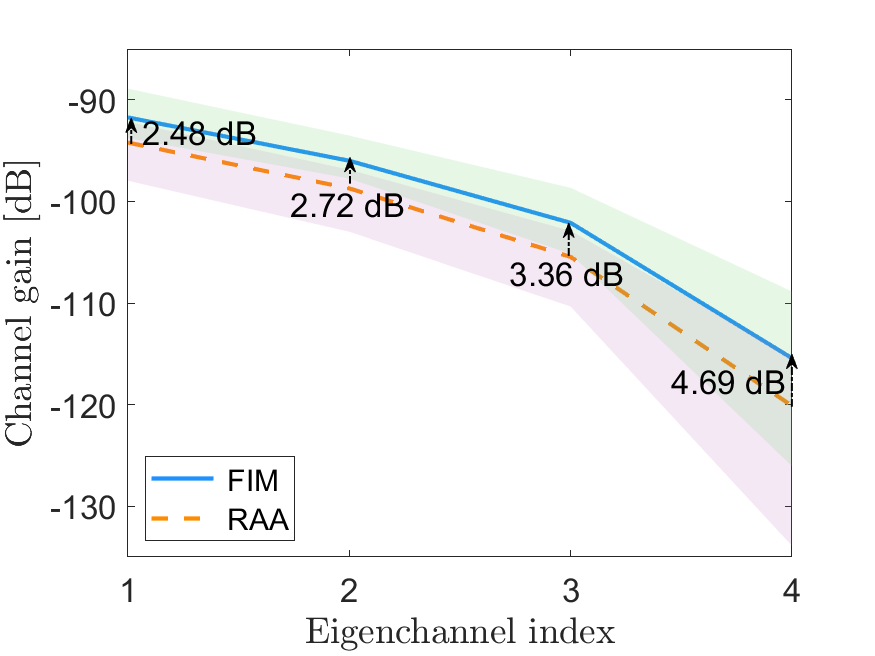}
\label{fig_10_1}}
\subfloat[$M = N = 16$.]{\includegraphics[width=4.25cm]{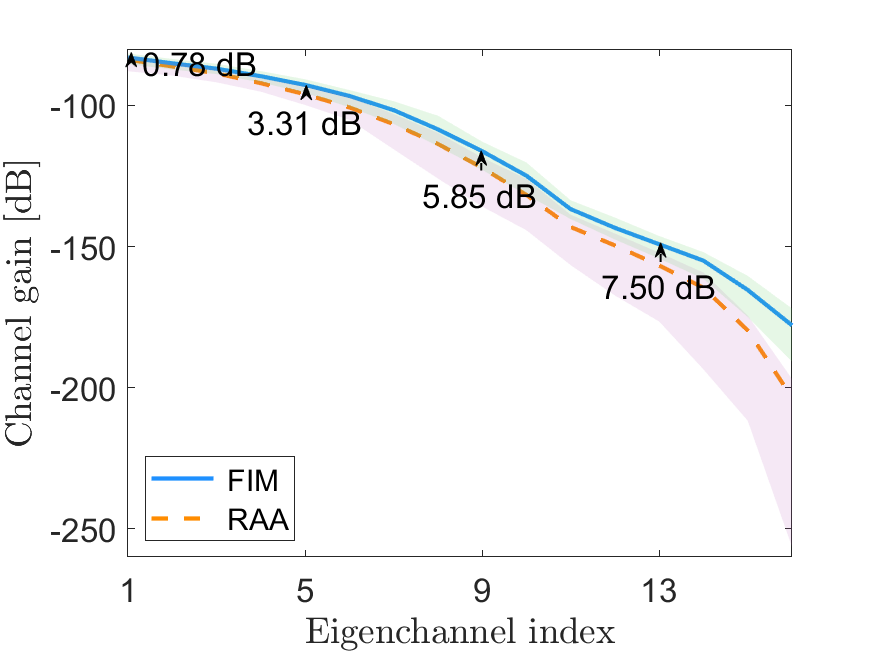}
\label{fig_10_2}}
\caption{Comparison of channel gains when utilizing RAAs and FIMs.}
\label{fig_10}
\end{figure}

\begin{figure}[!t]
\centering
\subfloat[$C$ versus the MSE of gain estimates.]{\includegraphics[width=4.25cm]{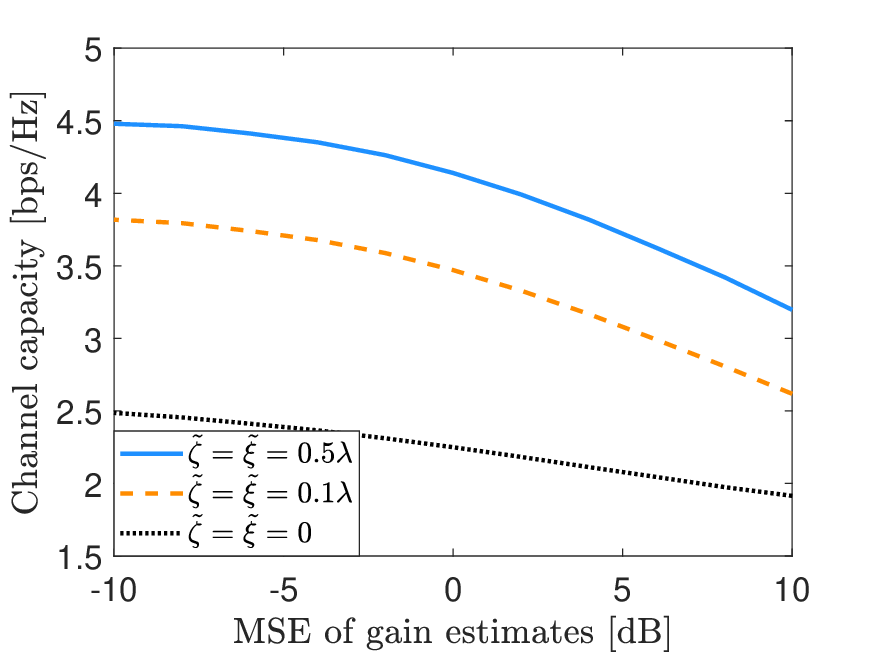}
\label{fig_10A_1}}
\subfloat[$C$ versus the RMSE of angle estimates.]{\includegraphics[width=4.25cm]{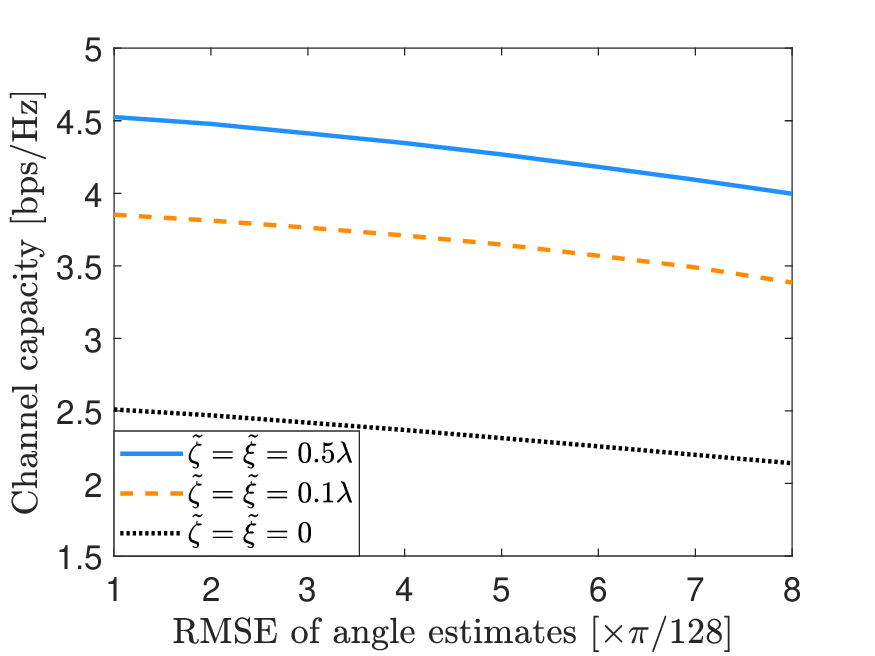}
\label{fig_10A_2}}
\caption{Channel capacity with imperfect CSI.}\vspace{-0.5cm}
\label{fig_10A}
\end{figure}
Next, Fig. \ref{fig_9} plots the channel capacity of different schemes as the morphing ranges of the FIMs increase. We consider $M = N = 4$ and keep all other parameters the same as in Fig. \ref{fig_3}. Note that a conventional RAA is a special case of an FIM when $\tilde \zeta = \tilde \xi = 0$. Observe from Fig. \ref{fig_9} that as $\tilde \zeta $ and $\tilde \xi$ increase, the transmitting and receiving FIMs exhibit increased flexibility to adapt their 3D surface shapes, thus gradually increasing the channel capacity. When the classic water-filling power allocation method is employed, the FIMs with a morphing range of $\tilde \zeta = \tilde \xi = 0.3\lambda$ can enhance the MIMO channel capacity from $2.5$ bps/Hz to $4.5$ bps/Hz. To put this into perspective, for a typical millimeter-wave frequency of $100$ GHz, this equates to a morphing range of just $\tilde \zeta = \tilde \xi = 1$ mm. While using the equal power allocation results in a moderate capacity penalty, the performance gain from morphing the surface shapes of FIMs is $2.4$ bps/Hz. Nevertheless, we observe diminishing returns as the morphing range continues to increase. Additionally, it is worth noting that the performance gains of FIMs are influenced by the normalized morphing range relative to the radio wavelength. For any given morphing range, the advantages of morphing FIM surface shapes become more pronounced as the operating frequency increases.

\begin{figure*}[!t]
\centering
\subfloat[]{\includegraphics[width=6cm]{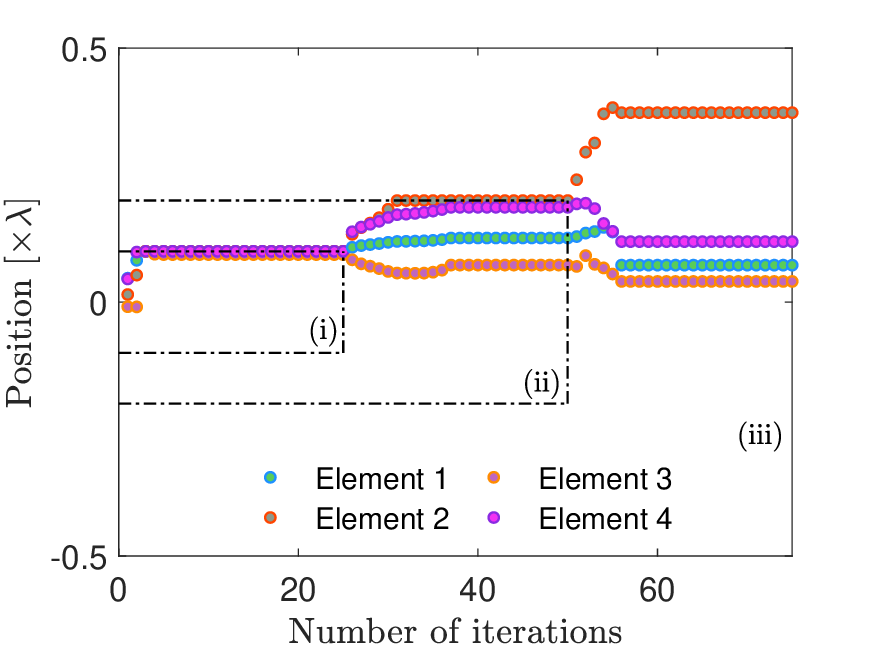}
\label{fig_11_1}}
\subfloat[]{\includegraphics[width=6cm]{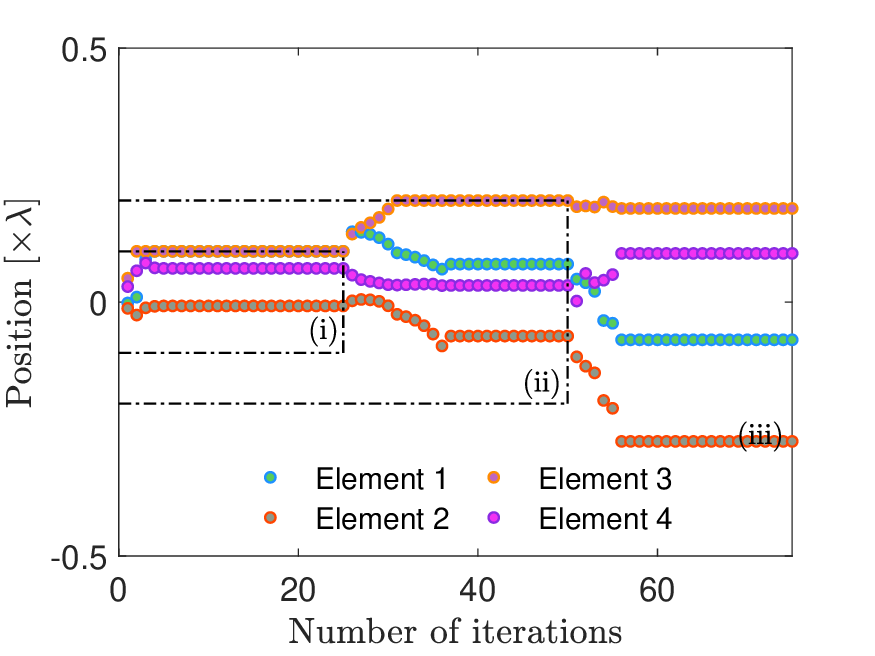}
\label{fig_11_2}}
\subfloat[]{\includegraphics[width=6cm]{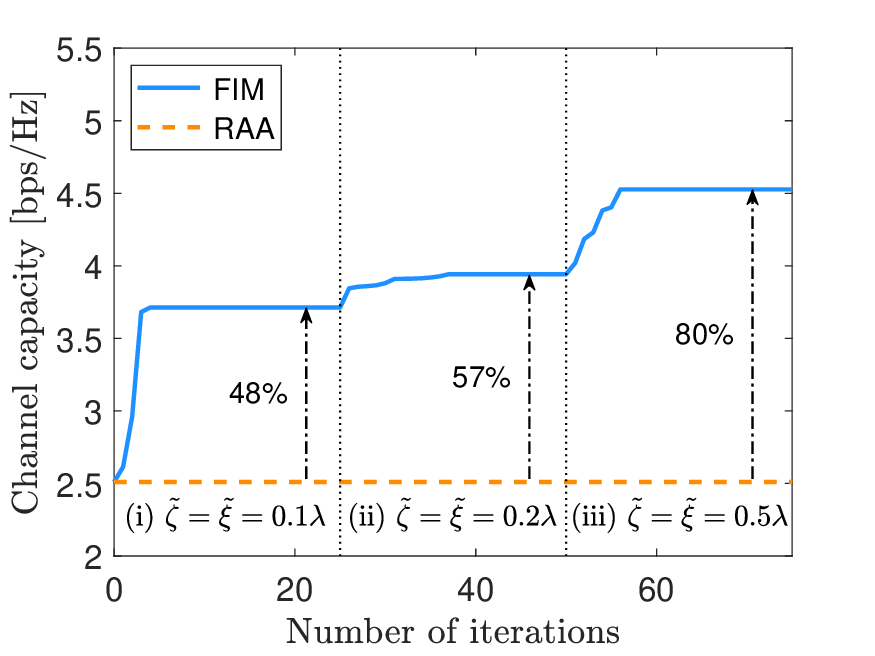}
\label{fig_11_3}}
\caption{Convergence behavior of (a) the transmitting FIM's surface shape; (b) the receiving FIM's surface shape; (c) the channel capacity, respectively, where we consider $L = 8$ scattering clusters.}
\label{fig_11}
\end{figure*}
\begin{figure*}[!t]
\centering
\subfloat[]{\includegraphics[width=6cm]{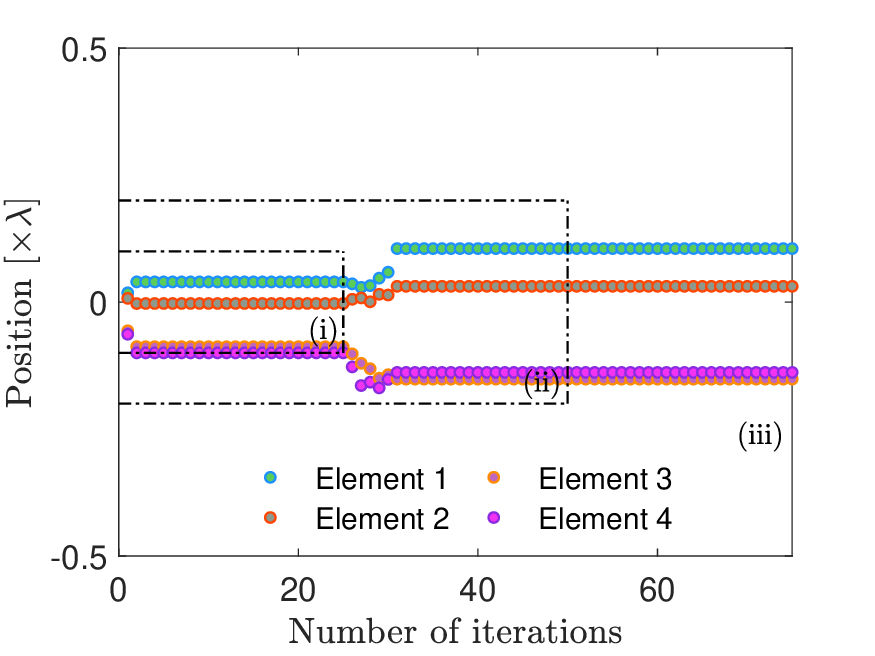}%
\label{fig_12_1}}
\subfloat[]{\includegraphics[width=6cm]{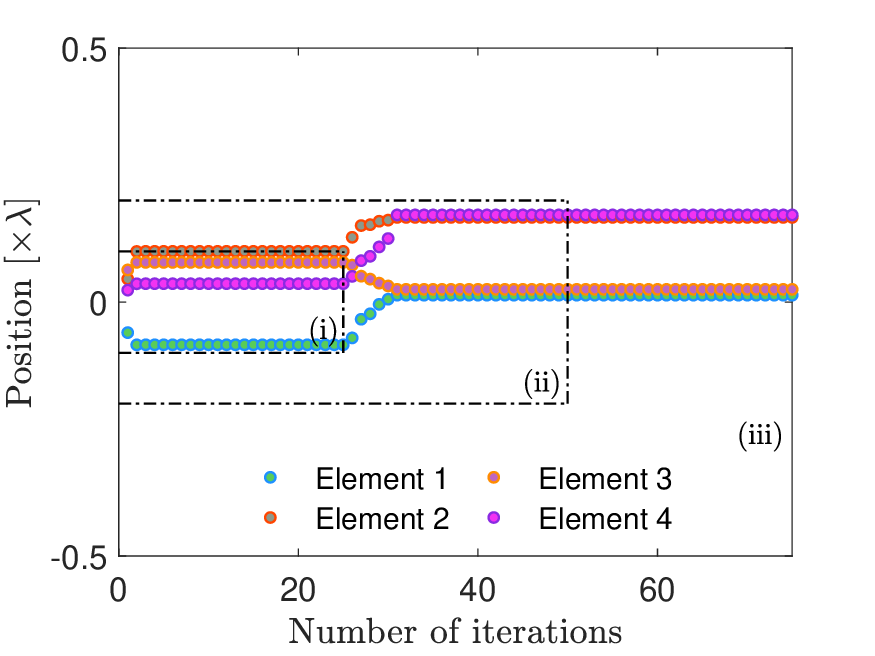}
\label{fig_12_2}}
\subfloat[]{\includegraphics[width=6cm]{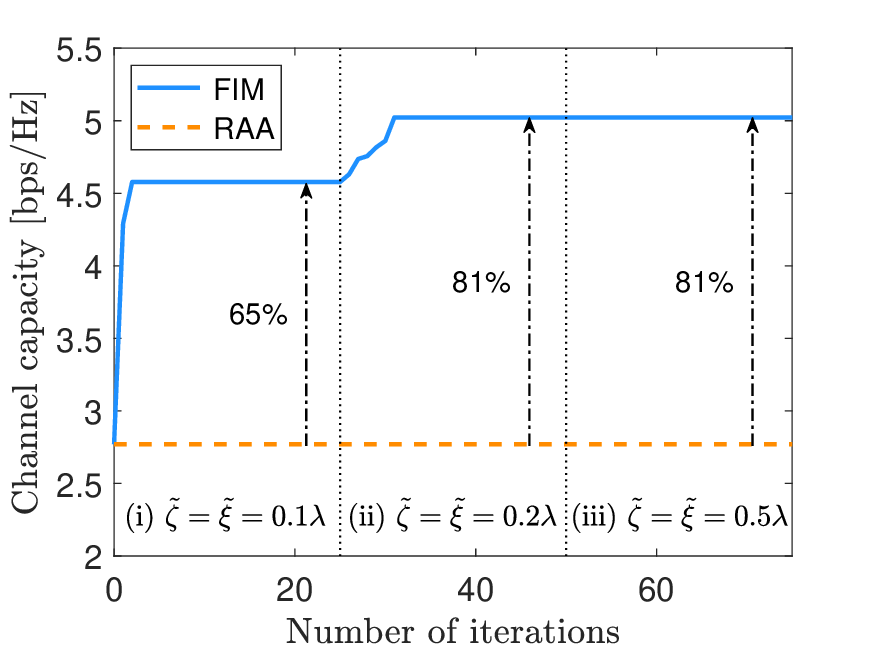}
\label{fig_12_3}}
\caption{Convergence behavior of (a) the transmitting FIM's surface shape; (b) the receiving FIM's surface shape; (c) the channel capacity, respectively, where we consider $L = 16$ scattering clusters.}
\label{fig_12}
\end{figure*}
\begin{figure*}[!t]
\centering
\subfloat[]{\includegraphics[width=6cm]{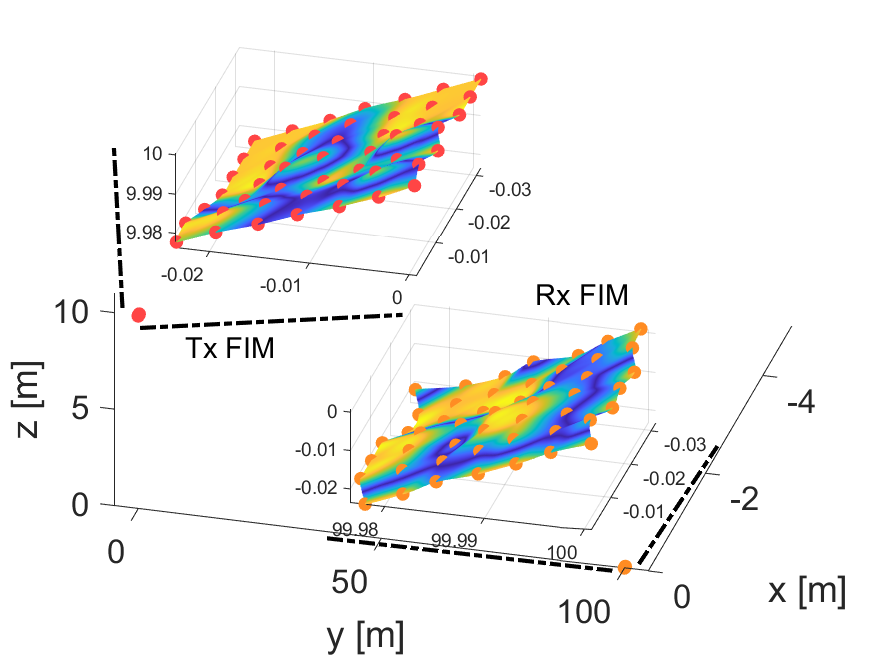}
\label{fig_13_1}}
\subfloat[]{\includegraphics[width=6cm]{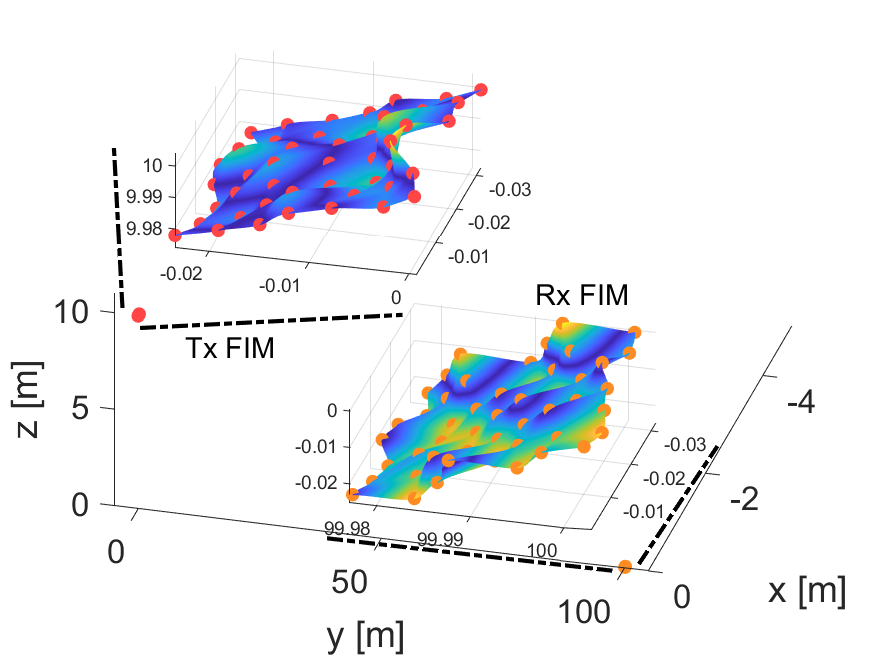}
\label{fig_13_2}}
\subfloat[]{\includegraphics[width=6cm]{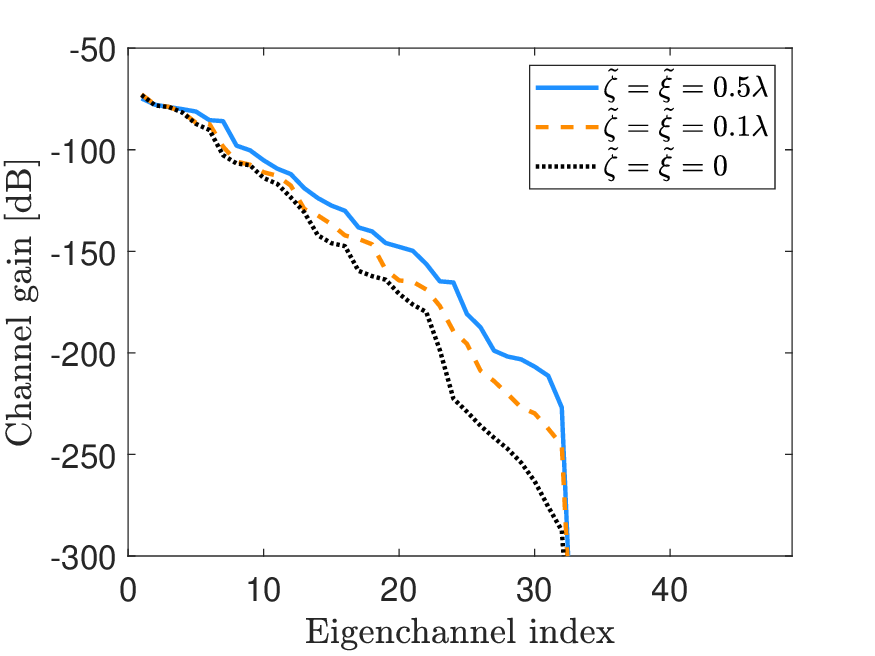}
\label{fig_13_3}}
\caption{(a) The morphed surface shapes of the transmitting and receiving FIMs, where we consider $\tilde \zeta = \tilde \xi = 0.1\lambda$; (b) The morphed surface shapes of the transmitting and receiving FIMs, where we consider $\tilde \zeta = \tilde \xi = 0.5\lambda$; (c) Channel gain under different values of the maximum morphing range.}\vspace{-0.5cm}
\label{fig_13}
\end{figure*}
\subsection{Channel Gain}\label{sec4_3}
To better illustrate the impact of the surface-shape morphing of the FIMs, we next evaluate the performance improvement in terms of eigenchannel gains. Fig. \ref{fig_10}(a) plots the average channel gain when utilizing $M = N = 4$ antenna elements for both transmission and reception, where the range of $100$ independent channel realizations is also shown in lighter colors. It is observed that FIMs are capable of enhancing the gains for all eigenchannels by adapting their surface shapes. Specifically, the average gains of the best and worst eigenchannels improve by $2.48$ dB and $4.69$ dB, respectively, indicating that the FIMs improve the quality of poorer channels more significantly. In Fig. \ref{fig_10}(b), we increase the number of transmitting and receiving antennas to $M = N = 16$. Again, square FIMs are considered. As shown in Fig. \ref{fig_10}(b), the FIMs improve the best eigenchannel gain by only $0.78$ dB, while improving the poorest eigenchannel by over $7.5$ dB. Additionally, compared to RAAs, the channel gain of utilizing FIMs is more stable across various channel environments, since the FIMs always strive to find the best 3D surface shapes for achieving high channel gain.

\subsection{Imperfect CSI}
Next, we examine the impact of imperfect CSI on channel capacity in Fig. \ref{fig_10A}, considering two different scenarios. In the first scenario (Fig. \ref{fig_10A}(a)), we focus on the imperfect estimate of the channel gain $\varsigma _{l,g},\ g=1,2,\ldots,G,\ l=1,2,\ldots,L$, which is modeled as a CSCG random variable, with a mean of $\varsigma_{l,g}$ and a variance corresponding to the MSE of the gain estimate. In the second scenario (Fig. \ref{fig_10A}(b)), we investigate the imperfect estimates of the azimuth and elevation angles. Specifically, the estimates for angles $\varphi_{l,g}^{\textrm{r}},\ \vartheta_{l,g}^{\textrm{r}},\ \varphi_{l,g}^{\textrm{t}},\ \vartheta_{l,g}^{\textrm{t}},\ g=1,2,\ldots,G,\ l=1,2,\ldots,L$ are modeled as uniform distributions centered around their corresponding actual values. For simplicity, we assume the same root MSE (RMSE) for all angle estimates, increasing it from $\pi/128$ to $\pi/16$. As shown in Fig. \ref{fig_10A}, imperfect estimates of both channel gain and angles lead to a degradation of MIMO channel capacity under all setups considered. Nonetheless, morphing the surface shapes of the transmitting and receiving FIMs consistently yields better channel capacity than conventional MIMOs relying on RAAs. Moreover, during the channel estimation phase, the surface shapes of FIMs can also be morphed to enhance channel estimation accuracy, while the specific protocol and algorithm design require further investigation.

\subsection{Convergence Analysis}\label{sec65}
Figs. \ref{fig_11} and \ref{fig_12} examine the convergence behavior of the proposed BCD algorithm. The simulation setup is the same as in Fig. \ref{fig_10}(a). In Fig. \ref{fig_11}, we consider a propagation environment having $L = 8$ scattering clusters. Specifically, Figs. \ref{fig_11}(a) and \ref{fig_11}(b) illustrate the position variations of the transmitting and receiving elements, respectively, as the iterations proceed, while the corresponding channel capacity is shown in Fig. \ref{fig_11}(c). Three cases are analyzed by progressively increasing the maximum morphing ranges of the transmitting and receiving FIMs:
\begin{enumerate}[label = (\roman*)]
 \item For iterations $\,\; 1$ -- $25$, we set $\tilde \zeta = \tilde \xi = 0.1 \lambda$;
 \item For iterations $26$ -- $50$, we set $\tilde \zeta = \tilde \xi = 0.2 \lambda$;
 \item For iterations $51$ -- $75$, we set $\tilde \zeta = \tilde \xi = 0.5\lambda$.
\end{enumerate}
As observed in Fig. \ref{fig_11}, the FIMs constantly morph their surface shapes to gradually increase the channel capacity, converging within $10$ iterations under all the setups considered. However, in each stage, the positions of some elements, e.g., transmitting element $2$ in Fig. \ref{fig_11}(a) and receiving element $3$ in Fig. \ref{fig_11}(b), are constrained by the maximum morphing range. Further increasing the maximum morphing ranges gives these elements more flexibility to adapt their positions. Interestingly, this also causes other elements to readjust their positions to cooperatively reconfigure the surface shapes of the transmitting and receiving FIMs. As a result, the channel capacity improves by $48$\%, $57$\%, and $80$\%, respectively, in the three setups compared to the conventional RAAs.

The results considering $L = 16$ scattering clusters are shown in Fig. \ref{fig_12}. In the first stage, the positions of transmitting element $4$ (shown in Fig. \ref{fig_12}(a)) and receiving element $2$ (shown in Fig. \ref{fig_12}(b)) are constrained by the maximum morphing range of $\tilde \zeta = \tilde \xi= 0.1\lambda$. Increasing the maximum morphing range to $0.2\lambda$ allows all elements to find their best positions. Therefore, the channel capacity is increased from $4.52$ bps/Hz to $4.96$ bps/Hz, as illustrated in Fig. \ref{fig_12}(c). Interestingly, as all elements already reach their optimal positions within the internal region of $\left ( -0.2\lambda, 0.2\lambda \right )$, further increasing the morphing range would no longer change the FIM surface shapes as well as the channel capacity. It is noted that in all setups, the proposed BCD method tends to converge within $10$ iterations. Again, comparing Figs. \ref{fig_12}(a) and \ref{fig_12}(b) concludes that the channel capacity improves as the number of scattering clusters increases.

\subsection{The Surface-Shape Morphing of FIMs}\label{sec63}
Finally, we demonstrate the capability of FIMs for morphing their surface shapes. Specifically, a pair of FIMs having $7 \times 7$ square antenna arrays are deployed at the source and destination, with all other parameters remaining the same as in Fig. \ref{fig_11}. Fig. \ref{fig_13}(a) shows the optimal surface shapes of the transmitting and receiving FIMs obtained by applying the proposed BCD method, when considering a maximum morphing range of $\tilde \zeta = \tilde \xi = 0.1\lambda$. It is demonstrated that the FIMs are capable of morphing their surface shapes by positioning all elements for improving the channel gain. As shown in Fig. \ref{fig_13}(c), weak eigenchannels are enhanced by over $40$ dB compared to conventional RAAs. In Fig. \ref {fig_13}(b), we increase the maximum morphing range of the transmitting and receiving FIMs to $\tilde \zeta = \tilde \xi = 0.5\lambda$. With a larger space to morph their surface shapes, the FIMs wiggle significantly more to enable the coherent superposition of multipath signal components. Consequently, the eigenchannel channel gain is further increased by about $20$ dB.

\section{Conclusions}\label{sec5}
In this paper, the FIM technology that has the capability of morphing its 3D surface shape was utilized as MIMO transceivers to enhance the channel capacity. Specifically, a capacity maximization problem was formulated and addressed by developing a customized BCD method for iteratively optimizing the transmit covariance matrix and the 3D surface shapes of the transmitting and receiving FIMs. Numerical results validated the significant capacity improvement gleaned from morphing the surface shapes of the FIMs compared to the conventional MIMO systems relying on RAAs. Notably, the performance gains become even more substantial as the maximum morphing range of the FIM and the number of propagation paths increase.

Additionally, integrating FIMs with BSs in wireless networks presents significant opportunities for enhancing multiuser communication performance. However, despite the remarkable performance gains achieved by dynamically morphing the 3D surface shapes of FIMs, further research is required to address challenges such as accurate channel estimation. It is also essential to consider the practical limitations of morphing FIM surface shapes and the associated energy consumption, which depends on several factors, including the substrate material, response time, and other practical factors. Furthermore, a system-level performance evaluation is necessary to characterize the fundamental trade-offs involved.

\appendices
\section{Proof of Proposition 1}\label{A1}
First, the channel capacity $C$ in \eqref{eq29} can be rewritten as
\begin{align}
 C = \log_{2}\det \mathbf{B}_{\textrm{r}}.
\end{align}

According to \emph{Eq. (3.60)} of \cite{TSP_2007_Hjorungnesa_Complex}, the differential of the logarithm of the determinant is given by
\begin{align}\label{eq52}
d C = \frac{1}{\ln 2}\textrm{tr}\left ( \mathbf{B}_{\textrm{r}}^{-1}d\mathbf{B}_{\textrm{r}} \right ).
\end{align}

Furthermore, we have
\begin{align}\label{eq53}
 d\mathbf{B}_{\textrm{r}}\overset{\left (a\right )}{=}&d\left [ \mathbf{A}_{\textrm{r}}\odot \mathbf{F}_{\textrm{r}}\left ( \boldsymbol{\xi} \right ) \right ]\mathbf{O}_{\textrm{t}}\left [ \mathbf{A}_{\textrm{r}}\odot \mathbf{F}_{\textrm{r}}\left ( \boldsymbol{\xi} \right ) \right ]^{H}\notag\\
&+\left [ \mathbf{A}_{\textrm{r}}\odot \mathbf{F}_{\textrm{r}}\left ( \boldsymbol{\xi} \right ) \right ]\mathbf{O}_{\textrm{t}}d\left [ \mathbf{A}_{\textrm{r}}\odot \mathbf{F}_{\textrm{r}}\left ( \boldsymbol{\xi} \right ) \right ]^{H}\notag\\
\overset{\left (b\right )}{=}&\left [ \mathbf{A}_{\textrm{r}}\odot d\mathbf{F}_{\textrm{r}}\left ( \boldsymbol{\xi} \right ) \right ]\mathbf{O}_{\textrm{t}}\left [ \mathbf{A}_{\textrm{r}}\odot \mathbf{F}_{\textrm{r}}\left ( \boldsymbol{\xi} \right ) \right ]^{H}\notag\\
&+\left [ \mathbf{A}_{\textrm{r}}\odot \mathbf{F}_{\textrm{r}}\left ( \boldsymbol{\xi} \right ) \right ]\mathbf{O}_{\textrm{t}}\left [ \mathbf{A}_{\textrm{r}}\odot d\mathbf{F}_{\textrm{r}}\left ( \boldsymbol{\xi} \right ) \right ]^{H},
\end{align}
where $\left (a\right )$ holds due to $d\left ( \mathbf{A}\mathbf{B} \right )=d\left ( \mathbf{A} \right )\mathbf{B}+\mathbf{A}d\left ( \mathbf{B} \right )$, while $\left ( b \right )$ holds due to $d\left ( \mathbf{A}\odot \mathbf{B} \right )=d\left ( \mathbf{A} \right )\odot \mathbf{B}+\mathbf{A}\odot d\left ( \mathbf{B} \right )$.

Hence, the partial derivative of the matrix $\mathbf{F}_{\textrm{r}}\left ( \boldsymbol{\xi} \right )$ \emph{w.r.t.} the deformation distance $\xi_{n} $ of the $n$-th receiving element is given by
\begin{align}\label{eq57}
 \frac{\partial }{\partial \xi_{n}}\mathbf{F}_{\textrm{r}}\left ( \boldsymbol{\xi}\right ) = j\tilde{\mathbf{F}}_{n}^{\textrm{r}}\left ( \boldsymbol{\xi} \right )\mathbf{K}_{\textrm{r}},\quad n=1,2,\ldots,N,
\end{align}
where $\tilde{\mathbf{F}}_{n}^{\textrm{r}}\left ( \boldsymbol{\xi} \right )\in \mathbb{C}^{N\times LG}$ is defined as
\begin{align}
 \tilde{\mathbf{F}}_{n}^{\textrm{r}}\left ( \boldsymbol{\xi} \right )= 
\begin{bmatrix}
\mathbf{0}_{\left ( n-1 \right )\times LG};\
\left [ \mathbf{F}_{\textrm{r}}\left ( \boldsymbol{\xi} \right ) \right ]_{n,:};\
\mathbf{0}_{\left ( N-n \right )\times LG}
\end{bmatrix}.
\end{align}

Substituting \eqref{eq57} into \eqref{eq53}, we arrive at
\begin{align}\label{eq58}
 \frac{\partial }{\partial \xi_{n}}\mathbf{B}_{\textrm{r}}=&\left ( \mathbf{A}_{\textrm{r}}\odot j\tilde{\mathbf{F}}_{\textrm{r}}\left ( \boldsymbol{\xi} \right )\mathbf{K}_{\textrm{r}} \right )\mathbf{O}_{\textrm{t}}\left ( \mathbf{A}_{\textrm{r}}\odot \mathbf{F}_{\textrm{r}}\left ( \boldsymbol{\xi} \right ) \right ) ^{H}\notag\\
&+\left ( \mathbf{A}_{\textrm{r}}\odot \mathbf{F}_{\textrm{r}}\left ( \boldsymbol{\xi} \right ) \right )\mathbf{O}_{\textrm{t}}\left ( \mathbf{A}_{\textrm{r}}\odot j\tilde{\mathbf{F}}_{\textrm{r}}\left ( \boldsymbol{\xi} \right )\mathbf{K}_{\textrm{r}} \right )^{H}\notag\\
=&-2\Im \left ( \tilde{\mathbf{S}}_{n}^{\textrm{r}} \right ),\quad n=1,2,\ldots,N,
\end{align}
where $\tilde{\mathbf{S}}_{n}^{\textrm{r}}\in \mathbb{C}^{N\times N}$ is defined as
\begin{align}
 \tilde{\mathbf{S}}_{n}^{\textrm{r}} = \begin{bmatrix}
 \boldsymbol{0}_{\left ( n-1 \right )\times \left ( n-1 \right )}& \mathbf{0}_{\left ( n-1 \right )\times 1} & \mathbf{0}_{\left ( n-1 \right )\times \left ( N-n \right )}\\ 
\mathbf{0}_{1\times \left ( n-1 \right )} & \left [ \mathbf{S}_{\textrm{r}} \right ]_{n,n} & \mathbf{0}_{1\times \left ( N-n \right )}\\ 
\mathbf{0}_{\left ( N-n \right )\times \left ( n-1 \right )} & \mathbf{0}_{\left ( N-n \right )\times 1} & \mathbf{0}_{\left ( N-n \right )\times \left ( N-n \right )}
\end{bmatrix}.
\end{align}

Substituting \eqref{eq58} into \eqref{eq52} and collecting $N$ partial derivatives into a matrix, the proof is completed.
\section{Proof of Proposition 2}\label{A2}
Rewriting the channel capacity as $C = \log_{2}\det \mathbf{B}_{\textrm{t}}$ and following the similar philosophy in Appendix \ref{A1}, the proof is completed.

\bibliographystyle{IEEEtran}
\bibliography{ref}

\begin{thebibliography}{10}
\providecommand{\url}[1]{#1}
\csname url@samestyle\endcsname
\providecommand{\newblock}{\relax}
\providecommand{\bibinfo}[2]{#2}
\providecommand{\BIBentrySTDinterwordspacing}{\spaceskip=0pt\relax}
\providecommand{\BIBentryALTinterwordstretchfactor}{4}
\providecommand{\BIBentryALTinterwordspacing}{\spaceskip=\fontdimen2\font plus
\BIBentryALTinterwordstretchfactor\fontdimen3\font minus \fontdimen4\font\relax}
\providecommand{\BIBforeignlanguage}[2]{{%
\expandafter\ifx\csname l@#1\endcsname\relax
\typeout{** WARNING: IEEEtran.bst: No hyphenation pattern has been}%
\typeout{** loaded for the language `#1'. Using the pattern for}%
\typeout{** the default language instead.}%
\else
\language=\csname l@#1\endcsname
\fi
#2}}
\providecommand{\BIBdecl}{\relax}
\BIBdecl

\bibitem{SCIS_2021_You_Towards}
X.~You, C.-X. Wang, J.~Huang, X.~Gao, Z.~Zhang, M.~Wang, Y.~Huang, C.~Zhang, Y.~Jiang, J.~Wang \emph{et~al.}, ``Towards {6G} wireless communication networks: Vision, enabling technologies, and new paradigm shifts,'' \emph{Sci. China Inf. Sci.}, vol.~64, no. 110301, pp. 1--74, Nov. 2021.

\bibitem{Proc_2021_Tataria_6G}
H.~Tataria, M.~Shafi, A.~F. Molisch, M.~Dohler, H.~Sjöland, and F.~Tufvesson, ``{6G} wireless systems: Vision, requirements, challenges, insights, and opportunities,'' \emph{Proc. IEEE}, vol. 109, no.~7, pp. 1166--1199, Jul. 2021.

\bibitem{JSAC_2020_Zhang_Prospective}
J.~Zhang, E.~Björnson, M.~Matthaiou, D.~W.~K. Ng, H.~Yang, and D.~J. Love, ``Prospective multiple antenna technologies for beyond {5G},'' \emph{IEEE J. Sel. Areas Commun.}, vol.~38, no.~8, pp. 1637--1660, Aug. 2020.

\bibitem{CM_2016_Bjornson_Massive}
E.~Björnson, E.~G. Larsson, and T.~L. Marzetta, ``Massive {MIMO}: ten myths and one critical question,'' \emph{IEEE Commun. Mag.}, vol.~54, no.~2, pp. 114--123, Feb. 2016.

\bibitem{arXiv_2023_An_Stacked_mag}
J.~An, C.~Yuen, C.~Xu, H.~Li, D.~W.~K. Ng, M.~Di~Renzo, M.~Debbah, and L.~Hanzo, ``Stacked intelligent metasurface-aided {MIMO} transceiver design,'' \emph{IEEE Wireless Commun.}, vol.~31, no.~4, pp. 123--131, Aug. 2024.

\bibitem{arXiv_2025_An_Emerging}
J.~An, M.~Debbah, T.~J. Cui, Z.~N. Chen, and C.~Yuen, ``Emerging technologies in intelligent metasurfaces: Shaping the future of wireless communications,'' \emph{arXiv preprint arXiv:2411.19754}, 2024.

\bibitem{JSAC_2003_Goldsmith_Capacity}
A.~Goldsmith, S.~Jafar, N.~Jindal, and S.~Vishwanath, ``Capacity limits of {MIMO} channels,'' \emph{IEEE J. Sel. Areas Commun.}, vol.~21, no.~5, pp. 684--702, Jun. 2003.

\bibitem{TIT_2003_Zheng_Diversity}
L.~Zheng and D.~Tse, ``Diversity and multiplexing: a fundamental tradeoff in multiple-antenna channels,'' \emph{IEEE Trans. Inf. Theory}, vol.~49, no.~5, pp. 1073--1096, May 2003.

\bibitem{TWC_2010_Marzetta_Noncooperative}
T.~L. Marzetta, ``Noncooperative cellular wireless with unlimited numbers of base station antennas,'' \emph{IEEE Trans. Wireless Commun.}, vol.~9, no.~11, pp. 3590--3600, Nov. 2010.

\bibitem{TWC_2017_Ngo_Cell}
H.~Q. Ngo, A.~Ashikhmin, H.~Yang, E.~G. Larsson, and T.~L. Marzetta, ``Cell-free massive {MIMO} versus small cells,'' \emph{IEEE Trans. Wireless Commun.}, vol.~16, no.~3, pp. 1834--1850, Mar. 2017.

\bibitem{TWC_2022_Lu_Communicating}
H.~Lu and Y.~Zeng, ``Communicating with extremely large-scale array/surface: Unified modeling and performance analysis,'' \emph{IEEE Trans. Wireless Commun.}, vol.~21, no.~6, pp. 4039--4053, Jun. 2022.

\bibitem{arXiv_2023_An_Toward}
J.~An, C.~Yuen, L.~Dai, M.~Di~Renzo, M.~Debbah, and L.~Hanzo, ``Near-field communications: Research advances, potential, and challenges,'' \emph{IEEE Wireless Commun.}, vol.~31, no.~3, pp. 100--107, Jun. 2024.

\bibitem{CM_2023_Cui_Near}
M.~Cui, Z.~Wu, Y.~Lu, X.~Wei, and L.~Dai, ``Near-field {MIMO} communications for {6G}: Fundamentals, challenges, potentials, and future directions,'' \emph{IEEE Commun. Mag.}, vol.~61, no.~1, pp. 40--46, Jan. 2023.

\bibitem{TCOM_2022_An_Low}
J.~An, C.~Xu, L.~Gan, and L.~Hanzo, ``Low-complexity channel estimation and passive beamforming for {RIS}-assisted {MIMO} systems relying on discrete phase shifts,'' \emph{IEEE Trans. Commun.}, vol.~70, no.~2, pp. 1245--1260, Feb. 2022.

\bibitem{LSA_2014_Cui_Coding}
T.~J. Cui, M.~Q. Qi, X.~Wan, J.~Zhao, and Q.~Cheng, ``Coding metamaterials, digital metamaterials and programmable metamaterials,'' \emph{Light: Science \& Applications}, vol.~3, no.~10, pp. e218--e218, Oct. 2014.

\bibitem{JSAC_2020_Renzo_Smart}
M.~Di~Renzo, A.~Zappone, M.~Debbah, M.-S. Alouini, C.~Yuen, J.~de~Rosny, and S.~Tretyakov, ``Smart radio environments empowered by reconfigurable intelligent surfaces: How it works, state of research, and the road ahead,'' \emph{IEEE J. Sel. Areas Commun.}, vol.~38, no.~11, pp. 2450--2525, Nov. 2020.

\bibitem{WC_2022_An_Codebook}
J.~An, C.~Xu, Q.~Wu, D.~W.~K. Ng, M.~D. Renzo, C.~Yuen, and L.~Hanzo, ``Codebook-based solutions for reconfigurable intelligent surfaces and their open challenges,'' \emph{IEEE Wireless Commun.}, vol.~31, no.~2, pp. 134--141, Apr. 2024.

\bibitem{WC_2021_Shlezinger_Dynamic}
N.~Shlezinger, G.~C. Alexandropoulos, M.~F. Imani, Y.~C. Eldar, and D.~R. Smith, ``Dynamic metasurface antennas for {6G} extreme massive {MIMO} communications,'' \emph{IEEE Wireless Commun.}, vol.~28, no.~2, pp. 106--113, Apr. 2021.

\bibitem{WC_2021_Deng_Reconfigurable}
R.~Deng, B.~Di, H.~Zhang, D.~Niyato, Z.~Han, H.~V. Poor, and L.~Song, ``Reconfigurable holographic surfaces for future wireless communications,'' \emph{IEEE Wireless Commun.}, vol.~28, no.~6, pp. 126--131, Dec. 2021.

\bibitem{CM_2022_Zhang_Intelligent}
H.~Zhang, S.~Zeng, B.~Di, Y.~Tan, M.~Di~Renzo, M.~Debbah, Z.~Han, H.~V. Poor, and L.~Song, ``Intelligent omni-surfaces for full-dimensional wireless communications: Principles, technology, and implementation,'' \emph{IEEE Commun. Mag.}, vol.~60, no.~2, pp. 39--45, Feb. 2022.

\bibitem{JSAC_2022_Zhang_Holographic}
H.~Zhang, H.~Zhang, B.~Di, M.~Di~Renzo, Z.~Han, H.~V. Poor, and L.~Song, ``Holographic integrated sensing and communication,'' \emph{IEEE J. Sel. Areas Commun.}, vol.~40, no.~7, pp. 2114--2130, Jul. 2022.

\bibitem{CM_2021_Dardari_Holographic}
D.~Dardari and N.~Decarli, ``Holographic communication using intelligent surfaces,'' \emph{IEEE Commun. Mag.}, vol.~59, no.~6, pp. 35--41, Jun. 2021.

\bibitem{CL_2023_An_A1}
J.~An, C.~Yuen, C.~Huang, M.~Debbah, H.~V. Poor, and L.~Hanzo, ``A tutorial on holographic {MIMO} communications—part {I}: Channel modeling and channel estimation,'' \emph{IEEE Commun. Lett.}, vol.~27, no.~7, pp. 1664--1668, Jul. 2023.

\bibitem{TWC_2022_Pizzo_Fourier}
A.~Pizzo, L.~Sanguinetti, and T.~L. Marzetta, ``Fourier plane-wave series expansion for holographic {MIMO} communications,'' \emph{IEEE Trans. Wireless Commun.}, vol.~21, no.~9, pp. 6890--6905, Sep. 2022.

\bibitem{CL_2023_An_A2}
J.~An, C.~Yuen, C.~Huang, M.~Debbah, H.~V. Poor, and L.~Hanzo, ``A tutorial on holographic {MIMO} communications—part {II}: Performance analysis and holographic beamforming,'' \emph{IEEE Commun. Lett.}, vol.~27, no.~7, pp. 1669--1673, Jul. 2023.

\bibitem{CL_2023_An_A3}
------, ``A tutorial on holographic {MIMO} communications—part {III}: Open opportunities and challenges,'' \emph{IEEE Commun. Lett.}, vol.~27, no.~7, pp. 1674--1678, Jul. 2023.

\bibitem{BOOK_2005_Tse_Fundamentals}
D.~Tse and P.~Viswanath, \emph{Fundamentals of Wireless Communication}.\hskip 1em plus 0.5em minus 0.4em\relax Cambridge University Press, 2005.

\bibitem{TSP_2018_Hu_Beyond}
S.~Hu, F.~Rusek, and O.~Edfors, ``Beyond massive {MIMO}: The potential of data transmission with large intelligent surfaces,'' \emph{IEEE Trans. Signal Process.}, vol.~66, no.~10, pp. 2746--2758, May 2018.

\bibitem{TWC_2020_Jung_Performance}
M.~Jung, W.~Saad, Y.~Jang, G.~Kong, and S.~Choi, ``Performance analysis of large intelligent surfaces ({LISs}): Asymptotic data rate and channel hardening effects,'' \emph{IEEE Trans. Wireless Commun.}, vol.~19, no.~3, pp. 2052--2065, Mar. 2020.

\bibitem{JSTSP_2016_Heath_An}
R.~W. Heath, N.~González-Prelcic, S.~Rangan, W.~Roh, and A.~M. Sayeed, ``An overview of signal processing techniques for millimeter wave {MIMO} systems,'' \emph{IEEE J. Sel. Topics Signal Process.}, vol.~10, no.~3, pp. 436--453, Apr. 2016.

\bibitem{TWC_2014_Ayach_Spatially}
O.~E. Ayach, S.~Rajagopal, S.~Abu-Surra, Z.~Pi, and R.~W. Heath, ``Spatially sparse precoding in millimeter wave {MIMO} systems,'' \emph{IEEE Trans. Wireless Commun.}, vol.~13, no.~3, pp. 1499--1513, Mar. 2014.

\bibitem{arXiv_2023_An_Stacked_DOA}
J.~An, C.~Yuen, Y.~L. Guan, M.~Di~Renzo, M.~Debbah, H.~Vincent~Poor, and L.~Hanzo, ``Two-dimensional direction-of-arrival estimation using stacked intelligent metasurfaces,'' \emph{IEEE J. Sel. Areas Commun.}, pp. 1--15, 2024, Early Access.

\bibitem{JSAC_2023_An_Stacked}
J.~An, C.~Xu, D.~W.~K. Ng, G.~C. Alexandropoulos, C.~Huang, C.~Yuen, and L.~Hanzo, ``Stacked intelligent metasurfaces for efficient holographic {MIMO} communications in {6G},'' \emph{IEEE J. Sel. Areas Commun.}, vol.~41, no.~8, pp. 2380--2396, Aug. 2023.

\bibitem{TWC_2022_Wong_Fluid}
K.-K. Wong and K.-F. Tong, ``Fluid antenna multiple access,'' \emph{IEEE Trans. Wireless Commun.}, vol.~21, no.~7, pp. 4801--4815, Jul. 2022.

\bibitem{TWC_2023_Ma_MIMO}
W.~Ma, L.~Zhu, and R.~Zhang, ``{MIMO} capacity characterization for movable antenna systems,'' \emph{IEEE Trans. Wireless Commun.}, vol.~23, no.~4, pp. 3392--3407, Apr. 2024.

\bibitem{JSTSP_2016_Sohrabi_Hybrid}
F.~Sohrabi and W.~Yu, ``Hybrid digital and analog beamforming design for large-scale antenna arrays,'' \emph{IEEE J. Sel. Topics Signal Process.}, vol.~10, no.~3, pp. 501--513, Apr. 2016.

\bibitem{WCL_2025_Huang_Stacked}
G.~Huang \emph{et~al.}, ``Stacked intelligent metasurfaces for task-oriented semantic communications,'' \emph{IEEE Wireless Commun. Lett.}, vol.~14, no.~2, pp. 310--314, Feb. 2025.

\bibitem{WCL_2024_Niu_Stacked}
H.~Niu \emph{et~al.}, ``Stacked intelligent metasurfaces for integrated sensing and communications,'' \emph{IEEE Wireless Commun. Lett.}, vol.~13, no.~10, pp. 2807--2811, Oct. 2024.

\bibitem{TCCN_2025_Hao_Multi}
\BIBentryALTinterwordspacing
H.~Liu \emph{et~al.}, ``Multi-user {MISO} with stacked intelligent metasurfaces: A {DRL}-based sum-rate optimization approach,'' 2024. [Online]. Available: \url{https://arxiv.org/abs/2408.04837}
\BIBentrySTDinterwordspacing

\bibitem{ICC_2024_An_Stacked}
J.~An, C.~Yuen, Y.~L. Guan, M.~Di~Renzo, M.~Debbah, H.~V. Poor, and L.~Hanzo, ``Stacked intelligent metasurface performs a {2D} {DFT} in the wave domain for {DOA} estimation,'' in \emph{Proc. IEEE Int. Conf. Commun. (ICC)}, 2024, pp. 3445--3451.

\bibitem{WCL_2024_Lin_Stacked}
S.~Lin \emph{et~al.}, ``Stacked intelligent metasurface enabled {LEO} satellite communications relying on statistical {CSI},'' \emph{IEEE Wireless Commun. Lett.}, vol.~13, no.~5, pp. 1295--1299, May 2024.

\bibitem{NE_2022_Liu_A}
C.~Liu, Q.~Ma, Z.~J. Luo, Q.~R. Hong, Q.~Xiao, H.~C. Zhang, L.~Miao, W.~M. Yu, Q.~Cheng, L.~Li \emph{et~al.}, ``A programmable diffractive deep neural network based on a digital-coding metasurface array,'' \emph{Nature Electronics}, vol.~5, no.~2, pp. 113--122, Feb. 2022.

\bibitem{arXiv_2023_An_Stacked}
\BIBentryALTinterwordspacing
J.~An, M.~Di~Renzo, M.~Debbah, H.~V. Poor, and C.~Yuen, ``Stacked intelligent metasurfaces for multiuser downlink beamforming in the wave domain,'' 2023. [Online]. Available: \url{https://arxiv.org/pdf/2309.02687}
\BIBentrySTDinterwordspacing

\bibitem{arXiv_2024_Liu_Stackedmag}
\BIBentryALTinterwordspacing
H.~Liu \emph{et~al.}, ``Stacked intelligent metasurfaces for wireless sensing and communication: Applications and challenges,'' 2024. [Online]. Available: \url{https://arxiv.org/pdf/2407.03566}
\BIBentrySTDinterwordspacing

\bibitem{Nature_2022_Bai_A}
Y.~Bai, H.~Wang, Y.~Xue, Y.~Pan, J.-T. Kim, X.~Ni, T.-L. Liu, Y.~Yang, M.~Han, Y.~Huang \emph{et~al.}, ``A dynamically reprogrammable surface with self-evolving shape morphing,'' \emph{Nature}, vol. 609, no. 7928, pp. 701--708, Sep. 2022.

\bibitem{NC_2016_Kamali_Decoupling}
S.~M. Kamali, A.~Arbabi, E.~Arbabi, Y.~Horie, and A.~Faraon, ``Decoupling optical function and geometrical form using conformal flexible dielectric metasurfaces,'' \emph{Nature Communications}, vol.~7, no. 11618, pp. 1--7, May 2016.

\bibitem{APSIS_2002_Georgakopoulos_HIRF}
S.~Georgakopoulos, C.~Birtcher, C.~Balanis, R.~Renaut, and A.~Panaretos, ``{HIRF} penetration and {PED} coupling analysis for fuselage models using a hybrid subgrid {FDTD(2,2)/FDTD(2,4)} method,'' in \emph{Proc. IEEE Antennas Propag. Society Int. Sym.}, vol.~2, 2002, pp. 690--693.

\bibitem{TAP_2020_Modi_Metasurface}
A.~Y. Modi, M.~A. Alyahya, C.~A. Balanis, and C.~R. Birtcher, ``Metasurface-based method for broadband {RCS} reduction of dihedral corner reflectors with multiple bounces,'' \emph{IEEE Trans. Antennas Propag.}, vol.~68, no.~3, pp. 1436--1447, Mar. 2020.

\bibitem{ISAP_2018_Modi_Investigation}
A.~Y. Modi, C.~A. Balanis, and C.~Birtcher, ``Investigation of checkerboard metasufaces on flexible curvilinear structure for {RCS} reduction,'' in \emph{Proc. IEEE Int. Sym. Antennas, Propag. \& USNC/URSI National Radio Science Meeting}, 2018, pp. 2329--2330.

\bibitem{ICC_2025_Jiancheng_Flexible}
\BIBentryALTinterwordspacing
J.~An, C.~Yuen, M.~Debbah, and L.~Hanzo, ``Flexible intelligent metasurfaces for enhanced {MIMO} communications,'' 2025. [Online]. Available: \url{https://arxiv.org/abs/2502.16478}
\BIBentrySTDinterwordspacing

\bibitem{GLOBECOM_2025_Jiancheng_Downlink}
\BIBentryALTinterwordspacing
J.~An, C.~Yuen, M.~D. Renzo, M.~Debbah, H.~V. Poor, and L.~Hanzo, ``Downlink multiuser communications relying on flexible intelligent metasurfaces,'' 2025. [Online]. Available: \url{https://arxiv.org/abs/2502.16472}
\BIBentrySTDinterwordspacing

\bibitem{TWC_2025_An_Flexible}
J.~An, C.~Yuen, M.~D. Renzo, M.~Debbah, H.~Vincent~Poor, and L.~Hanzo, ``Flexible intelligent metasurfaces for downlink multiuser {MISO} communications,'' \emph{IEEE Trans. Wireless Commun.}, pp. 1--14, 2025, Early Access.

\bibitem{TWC_2018_Shahmansoori_Position}
A.~Shahmansoori, G.~E. Garcia, G.~Destino, G.~Seco-Granados, and H.~Wymeersch, ``Position and orientation estimation through millimeter-wave {MIMO} in {5G} systems,'' \emph{IEEE Trans. Wireless Commun.}, vol.~17, no.~3, pp. 1822--1835, Mar. 2018.

\bibitem{TWC_2023_An_Fundamental}
J.~An, H.~Li, D.~W.~K. Ng, and C.~Yuen, ``Fundamental detection probability vs. achievable rate tradeoff in integrated sensing and communication systems,'' \emph{IEEE Trans. Wireless Commun.}, vol.~22, no.~12, pp. 9835--9853, Dec. 2023.

\bibitem{TCOM_2015_Rappaport_Wideband}
T.~S. Rappaport, G.~R. MacCartney, M.~K. Samimi, and S.~Sun, ``Wideband millimeter-wave propagation measurements and channel models for future wireless communication system design,'' \emph{IEEE Trans. Commun.}, vol.~63, no.~9, pp. 3029--3056, Sep. 2015.

\bibitem{TSP_2007_Hjorungnesa_Complex}
A.~Hjorungnes and D.~Gesbert, ``Complex-valued matrix differentiation: Techniques and key results,'' \emph{IEEE Trans. Signal Process.}, vol.~55, no.~6, pp. 2740--2746, Jun. 2007.

\end{thebibliography}

\end{document}